\documentclass[aps,pra,superscriptaddress,amsmath,amssymb,preprintnumbers,showpacs]{revtex4-2}
\usepackage{amssymb}
\usepackage{bm}
\usepackage{color}
\usepackage{filemod}

\def\ds{\displaystyle}
\def\sss{\scriptscriptstyle}

\begin{document}

Printed on \today, 
Last modified: \Filemodtoday{SpinBosonRWA_preprint.tex} 



\title{Quantum Dynamics, Master Equation and Equilibrium for a Qubit Coupled to a Thermal Boson Field}


\author{Hiromichi Nakazato}
\email{hiromici@waseda.jp --- ORCID:  0000-0002-5257-7309}
\affiliation{Department of Physics, Waseda University, Tokyo 169-8555, Japan}
\author{Saverio Pascazio}
\email{saverio.pascazio@uniba.it --- ORCID: 0000-0002-7214-5685}
\affiliation{Dipartimento di Fisica, Universit\`a di Bari, I-70126 Bari, Italy}
\affiliation{INFN, Sezione di Bari, I-70126 Bari, Italy}


\begin{abstract}
We analytically derive the exact---though formal---master equation for a two-level quantum system (qubit) interacting with a bosonic environment within the rotating-wave approximation, assuming the environment is initially in an arbitrary thermal state.
The long-time behavior of the evolution operator governing the dynamics of both the system and the environment is analyzed, and the conditions under which the system approaches thermal equilibrium are examined.

\end{abstract}

\maketitle

\section{Introduction}
Relaxation and thermalization are phenomena that occur universally in nature, making it essential to understand them within the framework of quantum theory---widely regarded as the fundamental theory of nature and one that has proven remarkably successful since its inception nearly a century ago.
Quantum theory's foundational equation---the Schr\"odinger equation (or equivalently, the von Neumann equation)---applies to isolated quantum systems. In such cases, no relaxation is expected, as the unitary evolution ensures that pure states remain pure indefinitely.
However, relaxation and thermalization are generally observed when a quantum system---typically a ``small" one---is coupled to a much larger external system or reservoir. To analyze these processes, one must consider the entire setup as a composite quantum system.
This involves treating the combined system (system plus reservoir) as a closed quantum system governed by the von Neumann equation for its total density matrix. To focus on the dynamics of the subsystem of interest, one derives a reduced density matrix by tracing out the reservoir's degrees of freedom.
Since this reduced density matrix captures the behavior of the open system, it serves as the foundation for understanding and deriving the relaxation and thermalization processes from a quantum mechanical perspective.

While the goal is conceptually straightforward, it is quite challenging to achieve in practice.
Solving the von Neumann equation to obtain the full time evolution of the total density matrix is nearly impossible due to the vast (often infinite) number of degrees of freedom involved.
As a result, a more practical approach is to derive directly a master equation that governs the dynamics of the small system alone. However, deriving this equation is itself nontrivial---it requires certain physical approximations, even though standard derivation techniques exist \cite{ME1,ME2,ME3,ME4,ME5,ME6,deVegaAlonso2017,Schlosshauer2007,BreuerPetruccione2007}.
Moreover, even after obtaining the master equation, one must still solve it to study relaxation and thermalization. This typically necessitates further approximations or, alternatively, numerical simulations.

In this context, exactly solvable models can offer valuable insights, although only a limited number of such models exist for open quantum systems \cite{deVegaAlonso2017,BMG,VB,ZLXTN,QBM,JTZY,BE}.
When the interaction between the system and the reservoir is of the dephasing type---meaning the interaction Hamiltonian commutes with the system Hamiltonian---the total system can be solved exactly, provided that the isolated system itself is analytically solvable (see, for example, Ref.\ \cite{WN2022}).
As the term suggests, dephasing interactions cause only decoherence in the system, without inducing transitions between energy levels.
To achieve thermalization or equilibrium, however, the interaction must be dissipative in nature, facilitating transitions between different system states.
Unfortunately, this requirement significantly complicates the search for solvable models that can capture thermalization dynamics and allow for analytical exploration of long-time behavior.
It is therefore quite remarkable that recent work has successfully demonstrated thermalization and analyzed asymptotic dynamics using solvable quantum models \cite{AliHuangZhang2020}, even though numerical simulations were still necessary to characterize the final equilibrium state.

In this paper, we investigate the well-known spin--boson model \cite{deVegaAlonso2017,Schlosshauer2007,BreuerPetruccione2007,BMG,VB,ZLXTN}, one of the few exactly solvable models for open quantum systems. We derive its master equation without employing any approximations and analyze its long-time behavior to gain insight into the relaxation and thermalization phenomena discussed earlier.
Our approach to deriving the master equation differs from the conventional methods \cite{deVegaAlonso2017,Schlosshauer2007,BreuerPetruccione2007}. Rather than beginning with perturbative expansions or assumptions about system-reservoir interactions, we first obtain the exact dynamical map---which evolves an initially uncorrelated system-reservoir state to a later-time state---in closed form. The master equation is then constructed directly from this exact solution.
This approach has an important advantage, in that it enable one to keep approximations and simplifications (such as those deriving from time-dependent coefficients) under control.

Importantly, we derive an exact master equation valid for any initial thermal state of the reservoir, provided there are no initial correlations between the system and the reservoir. This is in contrast to earlier studies \cite{BMG,VB,ZLXTN,SmirneVacchini2010,AliHuangZhang2020}, which focused only on special cases, such as when the reservoir begins in a vacuum state.
The resulting master equation takes the Gorini--Kossakowski--Lindblad--Sudarshan (GKLS) form \cite{GKLS1,GKLS2,CP}, featuring time-dependent coefficients. Crucially, it is time-local (or time-convolutionless), meaning it involves no time integrals, as long as the dynamical map remains invertible \cite{SmirneVacchini2010,ACH2007}. We emphasize that the invertibility of this map is intimately related to whether the system evolves toward equilibrium.

Since the time evolution operators for both the system and the reservoir are available in analytic form, we examine their asymptotic behavior. This allows us to study not only the long-time dynamics of the system but also those of the reservoir and to identify the conditions under which thermal equilibrium is established.
We find that, under reasonable assumptions, only on-shell contributions survive in the asymptotic operators. As a result, both the system and the reservoir relax to thermal states sharing the same temperature as the reservoir's initial thermal state.
This outcome aligns with the intuitive expectation that the system and the reservoir eventually equilibrate to thermal states characterized by the same temperature.

The structure of the paper is as follows.
In Section~\ref{s:model}, we introduce the spin--boson model. The time evolution of the reduced density matrix for a spin-1/2 system (qubit) is derived in Section~\ref{ss:rdm}, assuming the bosonic reservoir modes are initially in a thermal state.
Section~\ref{ss:EME} then constructs the master equation using the previously derived dynamical map. The resulting equation takes the GKLS form with time-dependent coefficients, expressed as expectation values of evolution operators over the initial thermal state of the reservoir.
In Section~\ref{s:asymptotic}, we investigate the system's asymptotic behavior and highlight the crucial role of the dynamical map's noninvertibility in determining whether equilibrium is reached. To analyze this, we examine the determinant of the transformation matrix, a quantity closely related to the reservoir's dynamics.
Section~\ref{ss:RD} presents an expression for the reservoir's reduced density matrix in terms of evolution operators that depend on the qubit states. These operators are shown to simplify significantly and can be expressed exactly using a set of transition-representing operators introduced earlier.
In Section~\ref{ss:Equilibrium}, we show that in the weak-coupling limit, the reservoir's reduced density matrix rapidly becomes stationary on a macroscopic time scale. At this stage, the relevant operators are composed of on-shell contributions that obey energy conservation. 
This highlights the presence of two distinct time scales: the first one is the time for the reservoir to go on-shell; the second (longer) one, is the evolution time typical of dissipation. 
The stationary state of the reservoir---corresponding to thermal equilibrium---is shown to match its initial thermal state, a consequence of unitarity relations. Finally, we demonstrate that the qubit system relaxes to a thermal state at the same temperature as the reservoir.
Section~\ref{s:summary} concludes the paper with a summary of results, discussion, and outlook for future work.
To make reading easier, a number of technical details have been omitted. Three appendices supplement the main text: Appendix~\ref{a:onshell} discusses the mechanism for realizing the asymptotic on-shell condition; Appendix~\ref{a:OL} elaborates on the asymptotic reservoir operators and shows that they are to be replaced by c-numbers; and Appendix~\ref{a:reservoir} addresses the special case of a zero-temperature (vacuum) reservoir.

\section{Model}
\label{s:model}
Consider a two-level quantum system (spin 1/2 or qubit) interacting with a bosonic environment (reservoir).  
The total system is assumed to be governed by the spin--boson Hamiltonian in the rotating-wave approximation (RWA)
\begin{equation}
H={\Omega\over2}\sigma_z+\sum_{\bm k}\bigl(\omega_{\bm k} a_{\bm k}^\dagger a_{\bm k}+g_{\bm k}(a_{\bm k}^\dagger\sigma_-+a_{\bm k}\sigma_+)\bigr)\equiv{\Omega\over2}\sigma_z+(\bm a^\dagger\bm\omega\bm a)+(\bm g\bm a^\dagger)\sigma_-+(\bm g\bm a)\sigma_+,
\label{eq:H}
\end{equation}
where the qubit is described by the Pauli matrices $\sigma_z,\sigma_\pm=(\sigma_x\pm i\sigma_y)/2$, $\Omega>0$ is the energy gap of the qubit, and $a_{\bm k}$ and $a_{\bm k}^\dagger$ stand for the annihilation and creation operators of the bosonic mode $\bm k$ with energy $\omega_{\bm k}$, satisfying the standard commutation relations, $[a_{\bm k},a^\dagger_{\bm k'}]=\delta_{\bm k,\bm k'}$, etc..  
In the right-hand side of Eq.\ (\ref{eq:H}) the summation over $\bm k$ is implicit in the scalar products among the quantities in boldface.
The real coupling function $g_{\bm k}$ describes the interaction between the qubit and mode $\bm k$. The RWA will enable the exact derivation of the evolution operator in a compact form and facilitate the discussion of its asymptotic behavior.

We first solve the von Neumann equation for the total system prepared in the initial state $\rho(0)e^{-\beta(\bm a^\dagger\bm\omega\bm a)}/{\cal Z}$, where the qubit is in an arbitrary initial state $\rho(0)$ and the bosonic modes are in the thermal state characterized by the inverse temperature $\beta$ with the normalization factor ${\cal Z}={\rm Tr}_a[e^{-\beta(\bm a^\dagger\bm\omega\bm a)}]$  (${\rm Tr}_a$ being the partial trace over the bosonic degrees of freedom).  
The reduced density matrix for the qubit in the interaction picture 
\begin{equation}
\rho_{\rm I}(t)={\rm Tr}_a\bigl[U(t)\rho(0)e^{-\beta(\bm a^\dagger\bm\omega\bm a)}U^\dagger(t)\bigr]/{\cal Z}
\label{eq:rhoI}
\end{equation}
is expressed in terms of the evolution operator in the interaction picture $U(t),\,U(0)=1$ ($\hbar=1$),
\begin{equation}
U(t)={\cal T}e^{-i\int_0^tdt_1H_{\rm I}(t_1)}=1-i\int_0^tdt_1H_{\rm I}(t_1)-\int_0^tdt_1\int_0^{t_1}dt_2H_{\rm I}(t_1)H_{\rm I}(t_2)+\cdots,
\end{equation}
where $\cal T$ stands for the time-ordering operator and the interaction Hamiltonian in the interaction picture is given by
\begin{equation}
 H_{\rm I}(t)=\sum_{\bm k}g_{\bm k}(e^{i(\omega_{\bm k}-\Omega)t}a_{\bm k}^\dagger\sigma_-+h.c.)\equiv(\bm g\bm a_t^\dagger)\sigma_-+(\bm g\bm a_t)\sigma_+.
\end{equation}
The evolution operator $U(t)$ acting on the total system is decomposed into four parts depending on the qubit states 
\begin{equation}
U(t)
=U_+(t)\sigma_+\sigma_--i\int_{0}^tdt'(\bm g\bm a_{t'}^\dagger)U_+(t')\sigma_-+U_-(t)\sigma_-\sigma_+-i\int_{0}^tdt'(\bm g\bm a_{t'})U_-(t')\sigma_+,
\label{eq:U(t)}
\end{equation}
where $U_\pm(t)$ are operators acting on the bosonic modes, when the qubit survives in the excited ($+$) or ground ($-$) states.
They explicitly read
\begin{align}
U_+(t)&=1-\int_0^tdt_1\int_0^{t_1}dt_2(\bm g\bm a_{t_1})(\bm g\bm a_{t_2}^\dagger)+\int_0^tdt_1\int_0^{t_1}dt_2\int_0^{t_2}dt_3\int_0^{t_3}dt_4(\bm g\bm a_{t_1})(\bm g\bm a_{t_2}^\dagger)(\bm g\bm a_{t_3})(\bm g\bm a_{t_4}^\dagger)+\cdots\nonumber\\
&\equiv1-\int_0^t(\bm g\bm a_{t_1})(\bm g\bm a_{t_2}^\dagger)+\int_0^t(\bm g\bm a_{t_1})(\bm g\bm a_{t_2}^\dagger)(\bm g\bm a_{t_3})(\bm g\bm a_{t_4}^\dagger)+\cdots
\label{eq:U+}
\end{align}
and
\begin{align}
U_-(t)&=1-\int_0^tdt_1\int_0^{t_1}dt_2(\bm g\bm a_{t_1}^\dagger)(\bm g\bm a_{t_2})+\int_0^tdt_1\int_0^{t_1}dt_2\int_0^{t_2}dt_3\int_0^{t_3}dt_4(\bm g\bm a_{t_1}^\dagger)(\bm g\bm a_{t_2})(\bm g\bm a_{t_3}^\dagger)(\bm g\bm a_{t_4})+\cdots\nonumber\\
&\equiv1-\int_0^t(\bm g\bm a_{t_1}^\dagger)(\bm g\bm a_{t_2})+\int_0^t(\bm g\bm a_{t_1}^\dagger)(\bm g\bm a_{t_2})(\bm g\bm a_{t_3}^\dagger)(\bm g\bm a_{t_4})+\cdots,
\label{eq:U-}
\end{align}
where the time-ordered integrals w.r.t.\ the variables indicated by the subscripts are implicit in the short-hand notation in the last lines. 

\subsection{Reduced Density Matrix at Time $t>0$}
\label{ss:rdm}
It is straightforward to extract the four components of the reduced density matrix $\rho_{\rm I}(t)$ by inserting the expression (\ref{eq:U(t)}) into (\ref{eq:rhoI}).  
In practice, the projection onto a particular qubit state and the partial trace over the bosonic degrees of freedom eliminate some non-relevant terms.
One can proceed as follows
\begin{align}
\sigma_+\rho_{\rm I}(t)\sigma_+&={\rm Tr}_a\Bigl[\bigl(-i\int_0^t(\bm g\bm a_{t_1}^\dagger)U_+(t_1)\sigma_+\sigma_-+U_-(t)\sigma_+\bigr)\rho(0)e^{-\beta(\bm a^\dagger\bm\omega\bm a)}\bigl(U_+^\dagger(t)\sigma_++i\int_0^tU_-^\dagger(t_1)(\bm g\bm a_{t_1}^\dagger)\sigma_-\sigma_+\bigr)\Bigr]/{\cal Z}\nonumber\\
&=\langle U_+^\dagger(t)U_-(t)\rangle\sigma_+\rho(0)\sigma_+,
\label{eq:U1}
\end{align}
where the brackets stand for the average over the initial bosonic state, $\langle\cdots\rangle={\rm Tr}_a[\cdots e^{-\beta(\bm a^\dagger\bm\omega\bm a)}]/{\cal Z}$.
Similarly, one obtains
\begin{align}
\label{eq:U2}
\sigma_-\rho_{\rm I}(t)\sigma_-
&=\langle U_-^\dagger(t)U_+(t)\rangle\sigma_-\rho(0)\sigma_-,\\
\sigma_+\rho_{\rm I}(t)\sigma_-
\label{eq:U3}
&=\Bigl\langle \int_0^tU_+^\dagger(t_1)(\bm g\bm a_{t_1})\int_0^t(\bm g\bm a_{t_2}^\dagger)U_+(t_2)\Bigr\rangle\sigma_+\sigma_-\rho(0)\sigma_+\sigma_-+\langle U_-^\dagger(t)U_-(t)\rangle\sigma_+\rho(0)\sigma_-,\\
\sigma_-\rho_{\rm I}(t)\sigma_+
\label{eq:U4}
&=\langle U_+^\dagger(t)U_+(t)\rangle\sigma_-\rho(0)\sigma_++\Bigl\langle\int_0^tU_-^\dagger(t_1)(\bm g\bm a_{t_1}^\dagger)\int_0^t(\bm g\bm a_{t_2})U_-(t_2)\Bigr\rangle\sigma_-\sigma_+\rho(0)\sigma_-\sigma_+.
\end{align}
We understand that while the first two components (\ref{eq:U1})--(\ref{eq:U2})
at time $t$ are directly related to those at initial time $t=0$,
\begin{equation}
\sigma_+\rho_{\rm I}(t)\sigma_+=\eta(t)\sigma_+\rho_{\rm I}(0)\sigma_+,\quad\sigma_-\rho_{\rm I}(t)\sigma_-=\eta^*(t)\sigma_-\rho_{\rm I}(0)\sigma_-,\quad\eta(t)\equiv\langle U_+^\dagger(t)U_-(t)\rangle,
\label{eq:++--}
\end{equation}
the remaining components (\ref{eq:U3})--(\ref{eq:U4}) are mixed and given by
\begin{equation}
\begin{pmatrix}
\sigma_+\rho_{\rm I}(t)\sigma_-\\\noalign{\smallskip}\sigma_+\sigma_-\rho_{\rm I}(t)\sigma_+\sigma_-
\end{pmatrix}=
\begin{pmatrix}
\alpha(t)&\zeta(t)\\\noalign{\smallskip}\gamma(t)&\xi(t)\end{pmatrix}\begin{pmatrix}
\sigma_+\rho(0)\sigma_-\\\noalign{\smallskip}\sigma_+\sigma_-\rho(0)\sigma_+\sigma_-
\end{pmatrix},
\label{eq:+-}
\end{equation}
and
\begin{equation}
\begin{pmatrix}
\sigma_-\rho_{\rm I}(t)\sigma_+\\\noalign{\smallskip}\sigma_-\sigma_+\rho_{\rm I}(t)\sigma_-\sigma_+
\end{pmatrix}=\begin{pmatrix}\xi(t)&\gamma(t)\\\noalign{\smallskip}\zeta(t)&\alpha(t)\end{pmatrix}\begin{pmatrix}\sigma_-\rho(0)\sigma_+\\\noalign{\smallskip}\sigma_-\sigma_+\rho(0)\sigma_-\sigma_+\end{pmatrix},
\label{eq:-+}
\end{equation}
where the nonnegative functions $\alpha(t),\gamma(t),\xi(t)$ and $\zeta(t)$ are defined as
\begin{align}
&\alpha(t)=\langle U_-^\dagger(t)U_-(t)\rangle,\quad\gamma(t)=\Bigl\langle\int_0^tU_-^\dagger(t_1)(\bm g\bm a_{t_1}^\dagger)\int_0^t(\bm g\bm a_{t_2})U_-(t_2)\Bigr\rangle,\\
&\xi(t)=\langle U_+^\dagger(t)U_+(t)\rangle,\quad\zeta(t)=\Bigl\langle\int_0^tU_+^\dagger(t_1)(\bm g\bm a_{t_1})\int_0^t(\bm g\bm a_{t_2}^\dagger)U_+(t_2)\Bigr\rangle.
\end{align}
These functions uniquely determine the reduced density matrix $\rho_{\rm I}(t)$ once the initial state $\rho(0)$ is given
\begin{align}
\rho_{\rm I}(t)&=(\sigma_+\sigma_-+\sigma_-\sigma_+)\rho_{\rm I}(t)(\sigma_+\sigma_-+\sigma_-\sigma_+)\nonumber\\
&=\gamma(t)\sigma_+\rho(0)\sigma_-+\xi(t)\sigma_+\sigma_-\rho(0)\sigma_+\sigma_-+\zeta(t)\sigma_-\rho(0)\sigma_++\alpha(t)\sigma_-\sigma_+\rho(0)\sigma_-\sigma_+\nonumber\\
&\quad+\eta^*(t)\sigma_+\sigma_-\rho(0)\sigma_-\sigma_++\eta(t)\sigma_-\sigma_+\rho(0)\sigma_+\sigma_-.
\label{eq:rhoI(t)}
\end{align}
This is the exact expression of the dynamical map, with the time dependence implicitly prescribed in the parameter functions. It is valid for any initial thermal state of the bosonic modes.  

Observe that the above reduced density matrix is properly normalized, i.e., probability is conserved, as it should be.  
Indeed,
\begin{equation}
{\rm Tr}[\rho_{\rm I}(t)]=(\alpha(t)+\gamma(t)){\rm Tr}[\sigma_-\sigma_+\rho(0)]+(\xi(t)+\zeta(t)){\rm Tr}[\sigma_+\sigma_-\rho(0)]={\rm Tr}[(\sigma_-\sigma_++\sigma_+\sigma_-)\rho(0)]=1,
\end{equation}
where the relations
\begin{equation}
\alpha(t)+\gamma(t)=\xi(t)+\zeta(t)=1
\label{eq:unity}
\end{equation}
have been used, representing the unitarity of the evolution operator $U^\dagger(t)U(t)=1$ in particular qubit states.  
Notice that the unitarity of $U(t)$ yields 
\begin{equation}
1=U_+^\dagger(t)U_+(t)+\int_0^tU_+^\dagger(t_1)(\bm g\bm a_{t_1})\int_0^t(\bm g\bm a_{t_2}^\dagger)U_+(t_2)=U_-^\dagger(t)U_-(t)+\int_0^tU_-^\dagger(t_1)(\bm g\bm a_{t_1}^\dagger)\int_0^t(\bm g\bm a_{t_2})U_-(t_2).
\label{eq:unitarity}
\end{equation}
These are operator relations in the bosonic space and their averages over the initial state are nothing but the relations (\ref{eq:unity}).  
Incidentally, the reduced density matrix can be expressed in the Kraus form \cite{Kraus,Kraus2} 
\begin{equation}
\rho_{\rm I}(t)=\sum_iK_i(t)\rho(0)K^\dagger_i(t),
\end{equation}
with, for example, 
\begin{equation}
K_i=\Bigl\{\sqrt{\gamma}\,\sigma_+,\sqrt{\zeta}\,\sigma_-,\sqrt\xi\Bigl(\sigma_+\sigma_-+{\eta\over\xi}\sigma_-\sigma_+\Bigr),\sqrt\alpha\sqrt{1-{|\eta|^2\over\alpha\xi}}\sigma_-\sigma_+\Bigr\},\qquad
\sum_iK_i^\dagger K_i=1.
\end{equation}
We note here that the (Schwarz) inequality $|\eta|^2\le\alpha\xi$ holds and all quantities in the square root are nonnegative.

\subsection{Exact Master Equation}
\label{ss:EME}
We now take an unusual route, and use the solution of the full dynamical problem 
(\ref{eq:H})--(\ref{eq:rhoI}) to derive the differential equation of the reduced dynamics of the system (master equation). 
It will become clear in the following that the very condition that such an equation can be constructed is closely related to the condition for the existence of an equilibrium.  

Given the reduced density matrix $\rho_{\rm I}(t)$ as an implicit function of time (\ref{eq:rhoI(t)}), it is simple to derive the master equation whose solution is $\rho_{\rm I}(t)$. 
We first take the time-derivative of $\rho_{\rm I}(t)$, obtaining it as a function of time and the initial state $\rho(0)$,   
and then reexpress $\rho(0)$ as a function of $\rho_{\rm I}(t)$ by inverting the relations (\ref{eq:++--})--(\ref{eq:-+}), to obtain the master equation. Explicitly:
\begin{align}
\dot\rho_{\rm I}(t)&=(\dot\gamma,\dot\xi)\begin{pmatrix}\sigma_+\rho(0)\sigma_-\\\noalign{\smallskip}\sigma_+\sigma_-\rho(0)\sigma_+\sigma_-\end{pmatrix}+(\dot\zeta,\dot\alpha)\begin{pmatrix}\sigma_-\rho(0)\sigma_+\\\noalign{\smallskip}\sigma_-\sigma_+\rho(0)\sigma_-\sigma_+\end{pmatrix}+\dot\eta^*\sigma_+\sigma_-\rho(0)\sigma_-\sigma_++\dot\eta\sigma_-\sigma_+\rho(0)\sigma_+\sigma_-\nonumber\\
&=(\dot\gamma,\dot\xi){1\over D}\begin{pmatrix}\xi&-\zeta\\\noalign{\smallskip}-\gamma&\alpha\end{pmatrix}\begin{pmatrix}\sigma_+\rho_{\rm I}(t)\sigma_-\\\noalign{\smallskip}\sigma_+\sigma_-\rho_{\rm I}(t)\sigma_+\sigma_-\end{pmatrix}+(\dot\zeta,\dot\alpha){1\over D}\begin{pmatrix}\alpha&-\gamma\\\noalign{\smallskip}-\zeta&\xi\end{pmatrix}\begin{pmatrix}\sigma_-\rho_{\rm I}(t)\sigma_+\\\noalign{\smallskip}\sigma_-\sigma_+\rho_{\rm I}(t)\sigma_-\sigma_+\end{pmatrix}\nonumber\\
&\quad+{\dot\eta^*\over\eta^*}\sigma_+\sigma_-\rho_{\rm I}(t)\sigma_-\sigma_++{\dot\eta\over\eta}\sigma_-\sigma_+\rho_{\rm I}(t)\sigma_+\sigma_-,
\label{eq:rhoIdot}
\end{align}
where 
\begin{equation}
D=\alpha\xi-\gamma\zeta=\alpha\xi-(1-\alpha)(1-\xi)=\alpha+\xi-1
\label{eq:ddef}
\end{equation}
is the determinant of the transformation matrices in (\ref{eq:+-}) and (\ref{eq:-+}).
After simple manipulations of the Pauli matrices, we can arrange the terms on the right-hand side of (\ref{eq:rhoIdot}) to obtain the exact master equation in the (time-dependent) GKLS form \cite{GKLS1,GKLS2,CP} 
\begin{align}
\dot\rho_{\rm I}(t)&=-i[G(t)\sigma_z,\rho_{\rm I}(t)]+\Gamma_z(t)\Bigl(\sigma_z\rho_{\rm I}(t)\sigma_z-\rho_{\rm I}(t)\Bigr)\nonumber\\
&\quad+\Gamma_-(t)\Bigl(\sigma_-\rho_{\rm I}(t)\sigma_+-{1\over2}\{\sigma_+\sigma_-,\rho_{\rm I}(t)\}\Bigr)+\Gamma_+(t)\Bigl(\sigma_+\rho_{\rm I}(t)\sigma_--{1\over2}\{\sigma_-\sigma_+,\rho_{\rm I}(t)\}\Bigr),\nonumber\\
\label{eq:ME}
\end{align}
where the coefficients $\Gamma_z,G$ and $\Gamma_\pm$ are all time dependent and read
\begin{align}
\Gamma_z&={\rm Re}F(t),\quad G=-{\rm Im}F(t),\\ 
F&=-{\dot\eta\over2\eta}+{\dot\xi+\dot\alpha\over4D}=-{1\over2}{d\over dt}\Bigl(\ln\eta-{1\over2}\ln(\alpha+\xi-1)\Bigr),\label{eq:Gamma_z}\\
\Gamma_-&={\alpha^2\over D}{d\over dt}\Bigl({1-\xi\over\alpha}\Bigr)=-{\alpha\dot\xi+(1-\xi)\dot\alpha\over\alpha+\xi-1},\label{eq:Gamma_-}\\
\Gamma_+&={\xi^2\over D}{d\over dt}\Bigl({1-\alpha\over\xi}\Bigr)=-{\dot\alpha\xi+(1-\alpha)\dot\xi\over\alpha+\xi-1}.
\label{eq:Gamma_+}
\end{align}
We stress that the above master equation has been derived without any approximation when the bosonic reservoir is in an arbitrary thermal state, for an arbitrary initial state of the qubit.  

This result is a natural extension of previous ones \cite{BMG,VB,ZLXTN,SmirneVacchini2010,AliHuangZhang2020} which focus on a reservoir that is initially in the vacuum state (zero-temperature limit of a thermal state), for a qubit initially in the excited state.  
We observe that the operator (GKLS) structure and the time dependence of the coefficients $\Gamma_\pm,\Gamma_z,G$
on the parameters ($\alpha,\xi,\eta$) are just the same as for the Jaynes--Cummings model \cite{SmirneVacchini2010}.  
When the bosonic mode of the Jaynes--Cummings model is generalized to a collection of modes in the spin--boson model, the strategy adopted for the former, that is, to relate matrix elements at different times, can no longer be applied owing to the degeneracies of the bosonic states, except for the particular case of a total excitation number equal to one.  

It should be stressed that we have obtained a time-local master equation, that does not involve time-integrals and memory effects.  
The strategy we adopted can in principle be applied to any discrete quantum system: we first decompose the reduced density matrix at time $t$ into independent operators, then relate them with those at the initial time $t=0$, take the derivative with respect to time, and finally invert the relations to get the master equation. 
Notice that a finite-dimensional matrix can be inverted unless its determinant vanishes, and the resulting equation is always a time-local one with time-dependent coefficients.  
In general, we always get time-local master equations for discrete quantum systems \cite{ACH2007,SmirneVacchini2010}, provided that the initial state is prepared in a product state (no initial correlations between system and environment) and the relation between operators at different times is invertible (in the present case, $D\not=0$).

\section{Asymptotic Behavior}
\label{s:asymptotic}

\subsection{Preliminary considerations}
\label{ss:prel}
Once the reduced density matrix is given as a function of time, a natural and interesting question arises ``What is the equilibrium state of the system, if it exists?".  
More fundamentally, one can ask ``Does the system always approach a particular stationary state at long times and what is the condition for the existence of such an equilibrium, that is independent of the initial state?". These are difficult questions, when one aims at full generality. However, the existence of an explicit solution and the implicit assumption that there is a unique equilibrium state enable us to write interesting formulas.

Let us examine the reduced density matrix at long times: this requires clarifying the asymptotic behavior of the relevant evolution operators, $U_\pm(t)$, etc..  
In particular, since the unique equilibrium state, if it exists, is independent of the initial state, we naturally expect that its existence must imply the loss of one-to-one correspondence with the initial state. In other words, the transformations (\ref{eq:+-}) and (\ref{eq:-+}) must become not invertible for the equilibrium state. 
We thus expect that the condition for the system to approach equilibrium is, from Eq.\ (\ref{eq:ddef}), 
\begin{equation}
D(t)\to0,\quad\hbox{i.e.,}\quad\alpha(t)+\xi(t)\to1\quad\hbox{as}\quad t\to\infty.
\end{equation}
The expectation appears reasonable.  
For example, since the probabilities of finding the system in the excited ($+$) or ground ($-$) states at time $t$ 
\begin{equation}
p_+(t)={\rm Tr}[\sigma_+\sigma_-\rho_{\rm I}(t)],\quad p_-(t)={\rm Tr}[\sigma_-\sigma_+\rho_{\rm I}(t)]
\end{equation}
are related to the initial probabilities by
\begin{equation}
\begin{pmatrix}p_+(t)\\p_-(t)\end{pmatrix}=\begin{pmatrix}\xi(t)&\gamma(t)\\\zeta(t)&\alpha(t)\end{pmatrix}\begin{pmatrix}p_+(0)\\p_-(0)\end{pmatrix}=\begin{pmatrix}\xi(t)\\\alpha(t)\end{pmatrix}+(1-\alpha(t)-\xi(t))\begin{pmatrix}0&1\\1&0\end{pmatrix}\begin{pmatrix}p_+(0)\\p_-(0)\end{pmatrix},    
\end{equation}
they can become independent of the initial conditions when the determinant vanishes $D=\alpha+\xi-1=0$.
If the last condition is satisfied at $t=\infty$, the state is to be characterized by the probabilities $p_+(\infty)=\xi(\infty)$ and $p_-(\infty)=\alpha(\infty)=1-p_+(\infty)$.

We are thus led to examine the condition $D\to0$ or $\alpha+\xi\to1$, which is a nontrivial and difficult problem, in general.  
Remember that the functions $\alpha$ and $\xi$ are expectation values, respectively of $U_-^\dagger U_-$ and $U_+^\dagger U_+$ over the initial thermal state, and the operators $U_\pm$ are defined via a time-ordered product and are iteratively written as in Eqs.\ (\ref{eq:U+}) and (\ref{eq:U-}).
Since the initial value $D(0)=\alpha(0)+\xi(0)-1=1$, and both $\alpha$ and $\xi \sim1+O(g^2)$, for weak coupling,  
one can (naively) wonder how the condition $D\to0$ can be attained in weak-coupling case, for $D\sim1+O(g^2)$.
As a matter of fact, even the master equation itself is usually derived under the assumption of a weakly coupled reservoir. The problem is clearly challenging. 
Our problem here is to examine the condition $D\to0$ and clarify, if possible, its relation to weak coupling.

Incidentally, probabilities become stationary when $\dot p_\pm\to0$, namely
\begin{equation}
0=\begin{pmatrix}\dot\xi&\dot\gamma\\\dot\zeta&\dot\alpha\end{pmatrix}{1\over D}\begin{pmatrix}\alpha&-\gamma\\-\zeta&\xi\end{pmatrix}\begin{pmatrix}p_+(t)\\p_-(t)\end{pmatrix}=\begin{pmatrix}-\Gamma_-&\Gamma_+\\\Gamma_-&-\Gamma_+\end{pmatrix}\begin{pmatrix}p_+(t)\\p_-(t)\end{pmatrix}.
\end{equation}
A stationary solution is possible when there exists $t_{\rm st}>0$ such that 
\begin{equation}
p_+(0)\dot\xi(t)=p_-(0)\dot\alpha(t), \quad \textrm{for all} \; t\ge t_{\rm st},
\end{equation}
so that probabilities become constant in time
\begin{align}
p_+^{\rm(st)}&={\Gamma_+(t)\over\Gamma_+(t)+\Gamma_-(t)}=\xi(t)+{\dot\xi(t)\over\dot\alpha(t)+\dot\xi(t)}(1-\alpha(t)-\xi(t))=\xi(t)+p_-(0)(1-\alpha(t)-\xi(t)),\\
p_-^{\rm(st)}&={\Gamma_-(t)\over\Gamma_+(t)+\Gamma_-(t)}=\alpha(t)+{\dot\alpha(t)\over\dot\alpha(t)+\dot\xi(t)}(1-\alpha(t)-\xi(t))=\alpha(t)+p_+(0)(1-\alpha(t)-\xi(t)).
\end{align}
It is clear that the above expressions yield the (initial-condition-independent) equilibrium probabilities when the condition $D=\alpha+\xi-1=0$ is met at $t=\infty$.

\subsection{Reservoir Dynamics}
\label{ss:RD}
We have seen that the condition for the qubit system to reach equilibrium is closely related to the noninvertibility of the dynamical map of the system and is given by $D\to0$ as $t\to\infty$.
The condition is expressed in terms of the evolution operators as $\langle U_+^\dagger U_+\rangle+\langle U_-^\dagger U_-\rangle\to1$ as $t\to\infty$.
It is in general hard to evaluate these expectation values and their limits.  
We notice that this condition is closely connected to the reduced density matrix of the bosonic modes $\rho_a$, as
\begin{equation} 
\langle U_\pm^\dagger(t)U_\pm(t)\rangle={\rm Tr}_a[U_\pm(t)\rho_a(0)U_\pm^\dagger(t)],\quad\rho_a(0)\equiv e^{-\beta(\bm a^\dagger\bm\omega\bm a)}/{\cal Z} .
\end{equation}
We thus try to scrutinize the features of $\rho_a(t)$, in order to gain insight into the present issue.  
We first recall that the reduced density matrix $\rho_a(t)$ is obtained from the total density matrix at time $t$ by tracing over the qubit degrees of freedom.  
Its time evolution depends on the initial and final qubit states. Explicitly,
\begin{align}
\rho_a(t)&=\Bigl[U_+(t)\rho_a(0)U_+^\dagger(t)+\int_0^t(\bm g\bm a_{t_1}^\dagger)U_+(t_1)\rho_a(0)\int_0^tU_+^\dagger(t_2)(\bm g\bm a_{t_2})\Bigl]\rho_{\sss++}(0)\nonumber\\
&\quad+\Bigl[U_-(t)\rho_a(0)U_-^\dagger(t)+\int_0^t(\bm g\bm a_{t_1})U_-(t_1)\rho_a(0)\int_0^tU_-^\dagger(t_2)(\bm g\bm a_{t_2}^\dagger)\Bigl]\rho_{\sss--}(0)\nonumber\\
&\quad+\Bigl[iU_+(t)\rho_a(0)\int_0^tU_-^\dagger(t_1)(\bm g\bm a_{t_1}^\dagger)-i\int_0^t(\bm g\bm a_{t_1}^\dagger)U_+(t_1)\rho_a(0)U_-^\dagger(t)\Bigl]\rho_{\sss+-}(0)+\hbox{\rm h.c.},
\label{eq:rhoat}
\end{align}
where $\rho_{\pm\pm}(0)$ are the four components of the initial density matrix of the qubit.  
In order to study the reservoir dynamics, we define an operator $A_t^\dagger$ by
\begin{equation}
\int_0^t(\bm g\bm a_{t'}^\dagger)U_+(t')\equiv A_t^\dagger U_+(t).
\label{eq:Adaggerdef}
\end{equation}
The time derivative of this equation yields
\begin{equation}
(\bm g\bm a_t^\dagger)U_+(t)=\dot{A}_t^\dagger U_+(t)-A_t^\dagger(\bm g\bm a_t)A_t^\dagger U_+(t), 
\end{equation}
which implies that $A_t^\dagger$ has to satisfy
\begin{equation}
\dot{A}_t^\dagger=(\bm g\bm a_t^\dagger)+A_t^\dagger(\bm g \bm a_t)A_t^\dagger,\quad A_0^\dagger=0.
\label{eq:Adaggerdot}
\end{equation}
The solution can be iteratively written
\begin{equation}
A_t^\dagger=\sum_{\bm k}(ga^\dagger)_{\bm k}{e^{i\bar\omega_{\bm k}t}-1\over i\bar\omega_{\bm k}}+\int_0^tA_{t'}^\dagger(\bm g\bm a_{t'})A_{t'}^\dagger,\quad\bar\omega_{\bm k}\equiv\omega_{\bm k}-\Omega.
\label{eq:Adaggert}
\end{equation}
It is not difficult to confirm that 
\begin{equation}
\int_0^t(\bm g\bm a_{t'})U_-(t')=A_tU_-(t),
\end{equation}
where $A_t$ has to satisfy conjugate of Eq.\ (\ref{eq:Adaggerdot})
\begin{equation}
\dot A_t=(\bm g\bm a_t)+A_t(\bm g\bm a_t^\dagger)A_t,\quad A_0=0.
\label{eq:Adot}
\end{equation}
The unitarity relations (\ref{eq:unitarity}), now written as $U_+^\dagger(t)U_+(t)+U_+^\dagger(t)A_tA_t^\dagger U_+(t)=1$ and $U_-^\dagger(t)U_-(t)+U_-^\dagger(t)A_t^\dagger A_tU_-(t)=1$, can be easily solved to yield the formal solutions for $U_\pm$
\begin{equation}
U_+(t)=(1+A_tA_t^\dagger)^{-{1\over2}}S_t,\quad U_-(t)=(1+A_t^\dagger A_t)^{-{1\over2}}V_t,
\label{eq:UV}
\end{equation}
where $S_t$ and $V_t$ are unitary operators. Notice that $A_tA_t^\dagger$ and $A_t^\dagger A_t$ are positive so that the above expressions are well defined.
These (exact!) expressions imply that the (dissipative) dynamics is solely governed by the operators $A_t$ and $A_t^\dagger$.
The parts of the reservoir reduced density matrix multiplied by $\rho_{++}(0)$ and $\rho_{--}(0)$ in Eq.\ (\ref{eq:rhoat}) are expressed, respectively, as
\begin{align}
\rho_{a+}(t)&=U_+(t)\rho_a(0)U_+^\dagger(t)+A_t^\dagger U_+(t)\rho_a(0)U_+^\dagger(t) A_t\nonumber\\
&={1\over\sqrt{1+A_tA_t^\dagger}}S_t\rho_a(0)S_t^\dagger{1\over\sqrt{1+A_tA_t^\dagger}}+A_t^\dagger{1\over\sqrt{1+A_tA_t^\dagger}}S_t\rho_a(0)S_t^\dagger{1\over\sqrt{1+A_tA_t^\dagger}}A_t\nonumber\\
&={1\over\sqrt{1+A_tA_t^\dagger}}S_t\rho_a(0)S_t^\dagger{1\over\sqrt{1+A_tA_t^\dagger}}+{1\over\sqrt{1+A_t^\dagger A_t}}A_t^\dagger S_t\rho_a(0)S_t^\dagger A_t{1\over\sqrt{1+A_t^\dagger A_t}}
\end{align}
and
\begin{align}
\rho_{a-}(t)&=U_-(t)\rho_a(0)U_-^\dagger(t)+A_tU_-(t)\rho_a(0)U_-^\dagger(t) A_t^\dagger\nonumber\\
&={1\over\sqrt{1+A_t^\dagger A_t}}V_t\rho_a(0)V_t^\dagger{1\over\sqrt{1+A_t^\dagger A_t}}+A_t{1\over\sqrt{1+A_t^\dagger A_t}}V_t\rho_a(0)V_t^\dagger{1\over\sqrt{1+A_t^\dagger A_t}}A_t^\dagger\nonumber\\
&={1\over\sqrt{1+A_t^\dagger A_t}}V_t\rho_a(0)V_t^\dagger{1\over\sqrt{1+A_t^\dagger A_t}}+{1\over\sqrt{1+A_tA_t^\dagger}}A_tV_t\rho_a(0)V_t^\dagger A_t^\dagger{1\over\sqrt{1+A_tA_t^\dagger}}.
\end{align}
These expressions are exact.

\subsection{Equilibrium}
\label{ss:Equilibrium}
We first show that the reservoir reduced density matrix $\rho_a(t)$ becomes stationary at relatively `short' times. 
Indeed, its time evolution is governed by
\begin{align}
\dot\rho_a(t)
&=\rho_{++}(0)\bigl(-[(\bm g\bm a_t),A_t^\dagger U_+(t)\rho_a(0)U_+^\dagger(t)]+[(\bm g\bm a_t^\dagger),U_+(t)\rho_a(0)U_+^\dagger(t)A_t]\bigl)\nonumber\\
&\quad+\rho_{--}(0)\bigl(-[(\bm g\bm a_t^\dagger),A_tU_-(t)\rho_a(0)U_-^\dagger(t)]+[(\bm g\bm a_t),U_-(t)\rho_a(0)U_-^\dagger(t)A_t^\dagger]\bigl)\nonumber\\
&\quad+i\rho_{+-}(0)\bigl(-[(\bm g\bm a_t),A_t^\dagger U_+(t)\rho_a(0)U_-^\dagger(t)A_t^\dagger]-[(\bm g\bm a_t^\dagger),U_+(t)\rho_a(0)U_-^\dagger(t)]\bigl)\nonumber\\
&\quad-i\rho_{-+}(0)\bigl([(\bm g\bm a_t^\dagger),A_tU_-(t)\rho_a(0)U_+^\dagger(t)A_t]+[(\bm g\bm a_t),U_-(t)\rho_a(0)U_+^\dagger(t)]\bigl).
\label{eq:rhoadot}
\end{align}  
Notice that the time-derivative $\dot\rho_a$ depends on commutators with $(\bm g\bm a_t)$ or $(\bm g\bm a_t^\dagger)$, and these operators behave, for example, like
\begin{equation}
(\bm g\bm a_t)=\sum_{\bm k}g_{\bm k}a_{\bm k}e^{-i\bar\omega_{\bm k}t}\to\sum_{\bm k,\bar\omega_{\bm k}=0}g_{\bm k}a_{\bm k}\quad\hbox{\rm for $t>t_R$},
\label{eq:tadef}
\end{equation}
where $t_R$ stands for a time scale over which the exponential factor $e^{-i\bar\omega_{\bm k}t}$ rapidly oscillates and yields no net contribution for any off-shell modes $\bm k$ with $\bar\omega_{\bm k}\not=0$.
The relevant time scales are detailed and discussed in Appendix~\ref{a:onshell}: $t_R$ is a characteristic time of the reservoir and is 
viewed as the time that it takes for the reservoir to ``forget". It is largely independent of the operators one considers (in this case $A_t$ and $A_t^\dagger$).

This is the point where our analysis becomes not exact and some (Fermi-Golden-rule-like) approximation is needed.
We can naively estimate the time scale $t_R$ as the inverse of the minimum energy-gap ($\hbar=1$) at resonance $\bar\omega_{\bm k}=\omega_{\bm k}-\Omega=0$, which is determined by the reservoir free dynamics and is considered to be very short compared with the relevant time scale for the qubit dynamics. This latter time scale is considered to be ``macroscopic", and is related to the approach to equilibrium of the qubit (see below). 
In addition, one needs a small coupling assumption. For this reason, we shall assume $g_{\bm k} \sim 1/N$, where $N$ is a scaling factor. In general, such a scaling factor is model-dependent. We can assume, in our case, that $N$ scales like the effective freedoms or the interaction volume.

We now endeavor to find the ``asymptotic" state of the reservoir.  
Notice that we are in a large time limit $t\gg t_R$ in terms of the reservoir (microscopic) dynamics and therefore only on-shell terms, that respect energy conservation, survive in the asymptotic reservoir density matrix. This somewhat intuitive statement requires justification. 
In this respect, we observe that the primitive functions $A_t$ and $A_t^\dagger$ that satisfy the differential equations (\ref{eq:Adaggerdot}) and (\ref{eq:Adot}) (irrespectively of the initial conditions) indeed preserve energy conservation at large times, for the exponential function $e^{ixt}/x$ appearing in the primitive functions becomes a delta function $e^{ixt}/x\to i\pi\delta(x)$ as $t\to\infty$. 
(A more careful treatment and additional technical details about how to achieve the on-shell condition are given in Appendix~\ref{a:onshell}, where this realization is justified by setting the initial condition at the remote past.) 
We obtain the asymptotic operators $A_t\to A$ and $A_t^\dagger\to A^\dagger$, which satisfy the commutation relations with the free Hamiltonian $(\bm a^\dagger\bm\omega\bm a)$
\begin{equation}
[A,(\bm a^\dagger\bm\omega\bm a)]=\Omega A,\quad[A^\dagger, (\bm a^\dagger\bm\omega\bm a)]=-\Omega A^\dagger.
\end{equation}
This is because each term in, say, $A$ contains a delta function $\delta(\bar\omega_{\bm k_1}-\bar\omega_{\bm k_2}+\bar\omega_{\bm k_3}-\cdots+\bar\omega_{\bm k_{2n+1}})$ that expresses energy conservation, multiplied by operator $a_{\bm k_1}a_{\bm k_2}^\dagger a_{\bm k_3}\cdots a_{\bm k_{2n+1}}$, its commutator with $(\bm a^\dagger\bm\omega\bm a)$ yields a factor $\omega_{\bm k_1}-\omega_{\bm k_2}+\omega_{\bm k_3}-\cdots+\omega_{\bm k_{2n+1}}$, which is replaced by the constant $\Omega$ owing to the presence of the delta function, and this factor can be pulled out as a common total factor.
A similar reasoning is applied to $A^\dagger$. 
The above relations imply that the asymptotic operators $AA^\dagger$ and $A^\dagger A$ commute with the free Hamiltonian $(\bm a^\dagger\bm\omega\bm a)$ and thus with $\rho_a(0)$.  
Here and in what follows $\rho_a(0)$ is taken to be the reservoir thermal state prepared at the remote past.
If we assume that the on-shell condition should also be (asymptotically) satisfied for the unitary operators $S_t$ and $V_t$ in Eq.\ (\ref{eq:UV}), these operators, consisting of pairs of creation and annihilation operators, become commutable with $\rho_a(0)$. The asymptotic counterparts of $\rho_{a\pm}(t)$ now read 
\begin{align}
\rho_{a+}^{\rm as}&={1\over\sqrt{1+AA^\dagger}}\rho_a(0){1\over\sqrt{1+AA^\dagger}}+A^\dagger{1\over\sqrt{1+AA^\dagger}}\rho_a(0){1\over\sqrt{1+AA^\dagger}}A={1\over1+AA^\dagger}\rho_a(0)+{1\over1+A^\dagger A}A^\dagger\rho_a(0)A\nonumber\\
&=\Bigl({1\over1+AA^\dagger}+{1\over1+A^\dagger A}A^\dagger Ae^{\beta\Omega}\Bigr)\rho_a(0),\\
\rho_{a-}^{\rm as}&={1\over\sqrt{1+A^\dagger A}}\rho_a(0){1\over\sqrt{1+A^\dagger A}}+A{1\over\sqrt{1+A^\dagger A}}\rho_a(0){1\over\sqrt{1+A^\dagger A}}A^\dagger={1\over1+A^\dagger A}\rho_a(0)+{1\over1+AA^\dagger}A\rho_a(0)A^\dagger\nonumber\\
&=\Bigl({1\over1+A^\dagger A}+{1\over1+AA^\dagger}AA^\dagger e^{-\beta\Omega}\Bigr)\rho_a(0).
\end{align}
We now remember that both $\rho_{a+}(t)$ and $\rho_{a-}(t)$ have been properly normalized to unity and both asymptotic operators $AA^\dagger$ and $A^\dagger A$ commute with $\rho_a(0)$. 
The normalization conditions for the asymptotic counterparts $\rho_{a\pm}^{\rm as}$ imply that operators $AA^\dagger$ and $A^\dagger A$ satisfy, under the trace over $\rho_a(0)$, 
\begin{equation}
{1\over1+AA^\dagger}+{1\over1+A^\dagger A}A^\dagger Ae^{\beta\Omega}=1={1\over1+A^\dagger A}+{1\over1+AA^\dagger}AA^\dagger e^{-\beta\Omega}.
\label{eq:operator}
\end{equation}
It is argued in Appendix~\ref{a:OL} that the above conditions can allow c-number solutions 
\begin{equation}
A^\dagger A={n\over\bar c-n},\quad AA^\dagger={n+1\over\bar c-(n+1)},\quad n\equiv{1\over e^{\beta\Omega}-1},
\label{eq:sol}
\end{equation}
where $\bar c>n+1$ is an arbitrary c-number.  
Remark that these solutions imply that the relations (\ref{eq:operator}) are satisfied as c-numbers and therefore the asymptotic reservoir state is nothing but the initial thermal state 
\begin{equation}
\rho_a^{\rm as}=\rho_{a+}^{\rm as}\cdot\rho_{++}(0)+\rho_{a-}^{\rm as}\cdot\rho_{--}(0)=\rho_a(0)\cdot\bigl(\rho_{++}(0)+\rho_{--}(0)\bigr)=\rho_a(0),
\end{equation}
only if the off-diagonal terms  asymptotically yield vanishing contributions. 
(The off-diagonal terms involve different---in the sense of their operator structure---unitary operators, say $S_t$ and $V_t^\dagger$, which would indefinitely oscillate for large $t$, yielding vanishing contributions; alternatively, one can assume that the weak-coupling limit, which essentially requires pairs of creation and annihilation operators, removes all off-diagonal terms.) 
We stress that the reservoir recovers the initial thermal state at microscopically long times, which, however, are considered to be short in a macroscopic sense.  
Indeed, the condition for the qubit to reach equilibrium $D=0$ is not met for an arbitrary $\bar c$, even though the reservoir has reached the thermal state for any $\bar c>n+1$.

Finally, we clarify the qubit asymptotic dynamics after the reservoir has reached equilibrium.  
As stressed above, we consider a time region that is microscopically very large, but macroscopically finite. In order to discuss the qubit asymptotic state and examine its equilibrium, we need to fix the value of $\bar c$.  
In this respect, observe that since the reservoir equilibrium state is irrelevant to the value of $\bar c$, we can even think of a time-dependent $\bar c$, where the time is macroscopic, much larger than $O(N)$, $t\gg t_R$.  
We seek a differential equation that $\bar c$ must satisfy. 
Recall that the solutions (\ref{eq:sol}) give the following expressions for $\alpha$ and $\xi$
\begin{equation}
\alpha=\langle U_-^\dagger U_-\rangle=\bigl\langle{1\over1+A^\dagger A}\bigr\rangle=1-{n\over\bar c},\quad
\xi=\langle U_+^\dagger U_+\rangle=\bigl\langle{1\over1+AA^\dagger}\bigr\rangle=1-{n+1\over\bar c}
\end{equation}
and that the coefficients $\Gamma_\pm$ in the master equation are related to the time derivatives of the quantities (\ref{eq:Gamma_-}) and (\ref{eq:Gamma_+}), which are now written as
\begin{align}
\Gamma_-&=-{\alpha^2\over D}{d\over dt}\bigl({D\over\alpha}\bigr)=\dot\alpha-\alpha{\dot D\over D}={n+1\over1-(2n+1)({1\over\bar c})}{d\over dt}\bigl({1\over\bar c}\bigr),\\
\Gamma_+&=-{\xi^2\over D}{d\over dt}\bigl({D\over\xi}\bigr)=\dot\xi-\alpha{\dot D\over D}={n\over1-(2n+1)({1\over\bar c})}{d\over dt}\bigl({1\over\bar c}\bigr).
\end{align}
These two expressions have to give the same value for ${1\over\bar c}$ and therefore we have $\Gamma_-(t)={n+1\over n}\Gamma_+(t)=e^{\beta\Omega}\Gamma_+(t)$, which implies that the qubit stationary state is the thermal state $\propto e^{-\beta\Omega\sigma_z/2}$. 
Since $\alpha=\xi=1$ at $t=0$ and the initial value of ${1\over\bar c}$ is zero, the solution is easily obtained
\begin{equation}
{1\over\bar c}={1\over2n+1}\Bigl[1-\exp\Bigl(-{2n+1\over n}\int_0^tdt'\Gamma_+(t')\Bigr)\Bigr]
\to{1\over2n+1} , \quad\hbox{\rm as}\quad t\to\infty,
\end{equation}
that is, the asymptotic value of $\bar c$ is $2n+1$. 
This also clarifies that the determinant $D$ is a monotonically decreasing function of $t$
\begin{equation}
D=\alpha+\xi-1=\exp\Bigl(-{2n+1\over n}\int_0^tdt'\Gamma_+(t')\Bigr)
\end{equation}
and vanishes only when $t\to\infty$. 
Notice that the above asymptotic value $\bar c=2n+1=\coth{\beta\Omega\over2}$ results in $A^\dagger A=e^{-\beta\Omega}$ and $AA^\dagger=e^{\beta\Omega}$. 
These results demonstrate that the qubit system reaches equilibrium at $t=\infty$, independently of the initial state, and the equilibrium is a thermal state with the same temperature as the (initial) reservoir state, which can also be confirmed by the asymptotic probabilities that the qubit is found in the excited ($+$) or ground ($-$) states
\begin{align}
p_+^{\rm as}&=\xi(\infty)={\rm Tr}_a\left[{1\over1+AA^\dagger}\rho_a(0)\right]={e^{-\beta\Omega/2}\over e^{-\beta\Omega/2}+e^{\beta\Omega/2}},\\
p_-^{\rm as}&=\alpha(\infty)={\rm Tr}_a\left[{1\over1+A^\dagger A}\rho_a(0)\right]={e^{\beta\Omega/2}\over e^{-\beta\Omega/2}+e^{\beta\Omega/2}}.
\end{align}

The above expressions for ${1\over\bar c}$ and $D$ illustrate that the relevant time scale for qubit relaxation is $\lambda^2 t$ with $\lambda$ characterizing the strength of coupling, for the coefficient $\Gamma_+$ is proportional to the squared coupling  $\propto\lambda^2$.  
The relevant time, scaling like $\lambda^{-2}\propto N^2$, can be considered as a macroscopic time scale: it is very large compared with the time scale $t_R$, that is relevant for the reservoir to reach equilibrium. 
This is a crucial indication of the relevance of the van Hove limit 
\cite{vanHove,Davies,Davies2,Palmer,SL,Ojima,ALV} for the system's thermalization process.

\section{Summary, Discussions, and Prospects}
\label{s:summary}

In this paper, we analyzed the dynamics of a composite quantum system consisting of a spin-1/2 particle (qubit) coupled to a collection of bosonic modes (reservoir), focusing on its long-time behavior and, in particular, its potential approach to equilibrium.
Although numerous studies have examined the same model and addressed similar questions, our work offers unique perspectives in several respects:
i) We begin by deriving the dynamical map for the qubit, which expresses its reduced density matrix at time 
$t>0$ in terms of its initial state at $t=0$. From this exact solution, the master equation is then derived. This approach contrasts with the conventional method, where the master equation is first formulated and then solved to obtain the dynamical map.
ii) Our exact master equation applies to any initial thermal state of the reservoir, extending beyond previous results that were limited to zero-temperature reservoirs \cite{BMG,VB,ZLXTN,SmirneVacchini2010,AliHuangZhang2020}.
iii) We consider the dynamics of both the qubit and the reservoir, enabling a detailed discussion of their asymptotic behavior. We explicitly demonstrate that the evolution of the qubit and the reservoir are interrelated through operators: the qubit's dynamics are governed by operators averaged or projected onto the reservoir's states, and vice versa.
iv) We clarify the asymptotic forms of the evolution operators based on unitarity relations. These operators can be simply constructed from the asymptotic operators $A$ and $A^\dagger$, which satisfy energy conservation.
v) We show that unitarity combined with the asymptotic on-shell condition imposes relations between the operators $AA^\dagger$ and $A^\dagger A$ under the trace with respect to the initial reservoir state $\rho_a(0)$ in Eq.\ (\ref{eq:operator}). This leads to a clear and intuitive understanding of the evolution of a small system interacting with a thermal reservoir.

As noted earlier, the exact master equation derived from the dynamical map is time-local (or time-convolutionless), involving no integrals over past times.
Our approach explicitly demonstrates that such a time-local master equation always exists for discrete quantum systems, provided the total system is initially in a product state (system and reservoir uncorrelated) and the dynamical map is invertible. This holds true even though deriving it can become much more complex under general interactions without the rotating wave approximation (RWA).
While this point has been acknowledged in prior works \cite{deVegaAlonso2017,ACH2007,SmirneVacchini2010}, explicit demonstrations remain rare.

We emphasize once again that the key quantity determining whether the system approaches equilibrium is 
$D$, the determinant of the transformation matrix in the dynamical map, which is directly connected to the evolution operator.
The value of $D$ indicates whether the dynamical map is invertible ($D\not=0$) or noninvertible ($D=0$). 
In the latter case, the system's state becomes independent of its initial state, signaling equilibration.
The behavior of $D$ can be analyzed through the solution for the parameter function $\bar c$. This analysis shows that $D$ is a monotonically decreasing function that tends toward zero as $t\to\infty$.

As noted earlier, in the long-time limit ($t=\infty$), both the qubit and the reservoir evolve under the influence of asymptotic operators and reach equilibrium. Notably, the reservoir rapidly returns to its initial thermal state.
This observation confirms a seemingly naive but physically plausible scenario: the reservoir, initially in thermal equilibrium, is momentarily perturbed by its interaction with the qubit but quickly relaxes back, while the qubit undergoes dissipation within a quasi-stationary environment.
This perspective is closely related to the framework of the Nakajima--Zwanzig projection method \cite{Nakajima1958,Zwanzig1960}, which treats open-system dynamics under similar assumptions.
The reservoir's rapid recovery is attributed to its free evolution as a macroscopic system in the weak-coupling limit, and this behavior is consistent with the so-called mixing property of thermal states \cite{Tasaki}.

It should be acknowledged, however, that the operator relations in Eq.~(\ref{eq:operator}) are nontrivial and play a crucial role---particularly in ensuring the normalization of the asymptotic reduced density matrices $\rho_{a\pm}^{\rm as}$.
Importantly, the uniqueness of these asymptotic solutions and their complete physical realization remain open issues and warrant further investigation (see the following discussion).
Nevertheless, in hindsight, we can say that these relations are sufficient to support a natural and intuitive picture of relaxation: a small system interacting with a thermal reservoir eventually thermalizes at the same temperature as the reservoir.
It is worth noting, however, that this conclusion does not apply universally. A significant exception occurs when the reservoir is initially in the vacuum or zero-temperature state---this special case is addressed separately in Appendix~\ref{a:reservoir}.

It is both instructive and important to recognize that, in order to achieve thermal equilibrium at large times ($t \to \infty$), the operators $A$ and $A^\dagger$ involved in the evolution must, in some sense, lose their operator character.
This is necessary because, in the asymptotic limit, expressions like $\rho_a(0)$, $A\rho_a(0)A^\dagger$ and $A^\dagger\rho_a(0)A$ must all correspond to the same equilibrium state. However, this equivalence is not possible if $A$ and $A^\dagger$ retain their full operator nature, as $\rho_a(0)$ cannot simultaneously be an eigenoperator of the superoperators  $A(\cdot)A^\dagger$ and $A^\dagger(\cdot)A$.  
Additionally, $A\rho_a(0)$ and $A^\dagger\rho_a(0)$ in Eq.\ (\ref{eq:rhoat}) should asymptotically vanish---an apparent contradiction with the persistence of the terms $A\rho_a(0)A^\dagger$ and $A^\dagger\rho_a(0)A$.
Clearly, such cases demand careful interpretation and appropriate mathematical treatment.

In this work, we have sidestepped the ambiguity of these limiting operators' behavior by instead relying on normalization conditions derived from their expectation values.
As shown in Appendix~\ref{a:OL}, the operator-based solutions do not satisfy the positivity requirement, which restricts us to solutions involving c-numbers.
Although these c-number solutions yield physically reasonable and expected results regarding relaxation, a systematic and explicit method for replacing operator expressions with state-dependent c-numbers remains an open problem.

Finally, we reiterate that the model exhibits two distinct time scales: one associated with the relaxation of the reservoir and the other with that of the qubit. The qubit's relaxation occurs on a macroscopic time scale, which scales as $\lambda^{-2}\propto N^2$ in the weak-coupling limit $\lambda\to0$ or $N\to\infty$.
Within this time scale, the reservoir relaxation time $t_R$ becomes negligibly small, even though it is large compared to the reservoir's intrinsic dynamics.

While this general picture of open quantum system dynamics has long been intuitively accepted and qualitatively described in the literature since the early days of quantum theory, rigorous analysis based on concrete, solvable models has remained scarce due to its inherent difficulty. The analysis and results presented in this paper aim to take a step toward addressing that challenge.

\section*{Acknowledgements} 
We are grateful to Berge Englert for valuable and stimulating discussions.
HN is partially supported by JSPS KAKENHI Grant-in-Aid for Scientific Research (C) 23K03268.
SP acknowledges support from PNRR MUR project PE0000023-NQSTI, from ``Budget MUR - Dipartimenti di Eccellenza 2023--2027''  - Quantum Sensing and Modelling for One-Health (QuaSiModO), and from INFN through the project ``QUANTUM''.

\appendix
\section{On-shell condition}
\label{a:onshell}
The operator $A_t^\dagger$ (as well as $A_t$) is explicitly (and exactly) given by (\ref{eq:Adaggert}), 
the first few terms of which read (mode summations implied)
\begin{align}
A_t^\dagger&=(ga^\dagger)_{\bm k}{e^{i\bar\omega_{\bm k}t}-1\over i\bar\omega_{\bm k}}\nonumber\\
&\quad
+{(ga^\dagger)_{\bm k_1}\over i\bar\omega_{\bm k_1}}(ga)_{\bm k_2}{(ga^\dagger)_{\bm k_3}\over i\bar\omega_{\bm k_3}}\Bigl({e^{i(\bar\omega_{\bm k_1}-\bar\omega_{\bm k_2}+\bar\omega_{\bm k_3})t}-1\over i(\bar\omega_{\bm k_1}-\bar\omega_{\bm k_2}+\bar\omega_{\bm k_3})}-{e^{i(\bar\omega_{\bm k_1}-\bar\omega_{\bm k_2})t}-1\over i(\bar\omega_{\bm k_1}-\bar\omega_{\bm k_2})}-{e^{i(-\bar\omega_{\bm k_2}+\bar\omega_{\bm k_3})t}-1\over i(-\bar\omega_{\bm k_2}+\bar\omega_{\bm k_3})}+{e^{-i\bar\omega_{\bm k_2}t}-1\over-i\bar\omega_{\bm k_2}}\Bigr)+\cdots.
\end{align}
The large $t$ limit is simply evaluated to be 
\begin{align}
A_t^\dagger&\to\pi(ga^\dagger)_{\bm k}\tilde\delta(\bar\omega_{\bm k})\nonumber\\
&\quad
+\pi{(ga^\dagger)_{\bm k_1}\over i\bar\omega_{\bm k_1}}(ga)_{\bm k_2}{(ga^\dagger)_{\bm k_3}\over i\bar\omega_{\bm k_3}}\Bigl(\tilde\delta(\bar\omega_{\bm k_1}-\bar\omega_{\bm k_2}+\bar\omega_{\bm k_3})-\tilde\delta(\bar\omega_{\bm k_1}-\bar\omega_{\bm k_2})-\tilde\delta(-\bar\omega_{\bm k_2}+\bar\omega_{\bm k_3})+\tilde\delta(\bar\omega_{\bm k_2})\Bigr)+\cdots,
\end{align}
where we have introduced 
\begin{equation}
\pi\tilde\delta(x)\equiv\pi\delta(x)-{\cal P}{1\over ix}
\end{equation}
with ${\cal P}$ denoting the principal part.
We observe that only a single term at each order in the coupling constant, represented by the delta functions $\delta(\bar\omega_{\bm k})$, $\delta(\bar\omega_{\bm k_1}-\bar\omega_{\bm k_2}+\bar\omega_{\bm k_3})$, etc., respects energy conservation and there remain off-shell contributions reflecting the initial condition at $t=0$, $A^\dagger_0=0$. 
(Incidentally, the primitive functions correspond to such single terms that preserve the energy.) 

Can we somehow get rid of such off-shell terms? Or alternatively, is the naive idea of asymptotic energy conservation, which seems physically natural, not valid in this model?  
At this point, notice that the definition of the bosonic creation/annihilation operators still leaves a possibility of introducing arbitrary degrees of freedom for their phase.  
We could have started with alternative operators with additional phases
\begin{equation}
a_{\bm k}\to a_{\bm k}e^{-i\bar\omega_{\bm k}t'},\quad a_{\bm k}^\dagger\to a_{\bm k}^\dagger e^{i\bar\omega_{\bm k}t'} ,
\end{equation}
with an arbitrary real parameter $t'$.  
These new operators satisfy the same canonical commutation relations and the initial thermal state $\rho_a(0)$ remains intact.  
The parameter $t'$ only appears in the interaction terms, $(ga)_{\bm k}\to(ga)_{\bm k}e^{-i\bar\omega_{\bm k}t'}$ and $(ga^\dagger)_{\bm k}\to(ga^\dagger)_{\bm k}e^{i\bar\omega_{\bm k}t'}$.   
The above replacement can be viewed as an additional free evolution by $t'$ in the interaction picture.  
Indeed it is equivalent to the replacement of the operators at $t=0$ with those at $t=t'$ and thus the initial condition, so far imposed at $t=0$, is now set at $t=-\infty$ when $t'\to-\infty$, which would eliminate the initial-condition dependent terms.  
Roughly speaking, this limit enforces energy conservation on the system due to rapidly oscillating phase factors for the energy-nonconserving terms.   
The expectation sounds plausible but it is nontrivial and for this reason we explicitly work out the limit in the following.

Assume that the asymptotic operators are defined according to the $t'=-\infty$ limit, for example,  
\begin{align}
A^\dagger=\lim_{t'\to-\infty}&\Bigl[\pi(ga^\dagger)_{\bm k}e^{i\bar\omega_{\bm k}t'}\tilde\delta(\bar\omega_{\bm k})\nonumber\\
&+\pi{\cal P}{(ga^\dagger)_{\bm k_1}\over i\bar\omega_{\bm k_1}}(ga)_{\bm k_2}{\cal P}{(ga^\dagger)_{\bm k_3}\over i\bar\omega_{\bm k_3}}e^{i(\bar\omega_{\bm k_1}-\bar\omega_{\bm k_2}+\bar\omega_{\bm k_3})t'}\nonumber\\
&\qquad
\times\Bigl(\tilde\delta(\bar\omega_{\bm k_1}-\bar\omega_{\bm k_2}+\bar\omega_{\bm k_3})-\tilde\delta(\bar\omega_{\bm k_1}-\bar\omega_{\bm k_2})-\tilde\delta(-\bar\omega_{\bm k_2}+\bar\omega_{\bm k_3})+\tilde\delta(\bar\omega_{\bm k_2})\Bigr)+\cdots\Bigr],
\end{align}
where the symbol $\cal P$ has been explicitly inserted, for there are no singularities in the above expression and actually the numerator vanishes when the denominator vanishes.  
We notice that the limit ${\cal P}e^{ixt'}/ix\to-\pi\delta(x)$ exists as $t'\to-\infty$ and thus $\pi\tilde\delta(x)e^{ixt'}\to2\pi\delta(x)$.  
We therefore have
\begin{align}
A^\dagger&=2\pi(ga^\dagger)_{\bm k}\delta(\bar\omega_{\bm k})\nonumber\\
&\quad
+2\pi(ga^\dagger)_{\bm k_1}(ga)_{\bm k_2}(ga^\dagger)_{\bm k_3}\delta(\bar\omega_{\bm k_1}-\bar\omega_{\bm k_2}+\bar\omega_{\bm k_3}){\cal P}\Bigl({1\over i\bar\omega_{\bm k_1}i\bar\omega_{\bm k_3}}+{\pi\delta(\bar\omega_{\bm k_3})\over i\bar\omega_{\bm k_1}}+{\pi\delta(\bar\omega_{\bm k_1})\over i\bar\omega_{\bm k_3}}+\pi^2\delta(\bar\omega_{\bm k_1})\delta(\bar\omega_{\bm k_3})\Bigr)+\cdots\nonumber\\
&=2\pi(ga^\dagger)_{\bm k}\delta(\bar\omega_{\bm k})+2\pi{(ga^\dagger)_{\bm k_1}\over i\bar\omega_{\bm k_1}+\epsilon}(ga)_{\bm k_2}{(ga^\dagger)_{\bm k_3}\over i\bar\omega_{\bm k_3}+\epsilon}\delta(\bar\omega_{\bm k_1}-\bar\omega_{\bm k_2}+\bar\omega_{\bm k_3})\nonumber\\
&\quad
+2\pi{(ga^\dagger)_{\bm k_1}\over i\bar\omega_{\bm k_1}+\epsilon}(ga)_{\bm k_2}{(ga^\dagger)_{\bm k_3}\over i\bar\omega_{\bm k_3}+\epsilon}(ga)_{\bm k_4}{(ga^\dagger)_{\bm k_5}\over i\bar\omega_{\bm k_5}+\epsilon}\delta(\bar\omega_{\bm k_1}-\bar\omega_{\bm k_2}+\bar\omega_{\bm k_3}-\bar\omega_{\bm k_4}+\bar\omega_{\bm k_5})\nonumber\\
&\qquad
\times\Bigl({1\over i(\bar\omega_{\bm k_1}-\bar\omega_{\bm k_2}+\bar\omega_{\bm k_3})+\epsilon}+{1\over i(\bar\omega_{\bm k_3}-\bar\omega_{\bm k_4}+\bar\omega_{\bm k_5})+\epsilon}\Bigr)+\cdots,
\label{eq:Adagger}
\end{align}
where $\epsilon>0$ is an infinitesimal quantity and the next fifth-order term, so far suppressed, has been explicitly worked out.
We easily understand that energy conservation has been recovered in $A^\dagger$ in the form of delta functions, $\delta(\bar\omega_{\bm k}), \delta(\bar\omega_{\bm k_1}-\bar\omega_{\bm k_2}+\bar\omega_{\bm k_3}),\delta(\bar\omega_{\bm k_1}-\bar\omega_{\bm k_2}+\bar\omega_{\bm k_3}-\bar\omega_{\bm k_4}+\bar\omega_{\bm k_5})$, etc.\ at each (odd) order in coupling function.  
Needless to say, the hermitian conjugate of the above expression yields the asymptotic operator $A$ and therefore $A$ is composed of only on-shell terms as well. 
The same results can be obtained if one starts, from the outset, with initial condition at $t'$ and takes the limits $t\to\infty$ and $t'\to-\infty$. 
This concludes our proof.

We stress that, in the above expressions, ``large'' $t$ is relative to the free dynamics of the reservoir and one can replace the delta functions with the time duration $t \, (\sim2 \pi \delta (0))$, which is interpreted as the relaxation time of the reservoir. This enables one to replace, for instance,
\begin{align}
 2\pi(ga^\dagger)_{\bm k}\delta(\bar\omega_{\bm k})\to (ga^\dagger)_{\omega_{\bm k}=\Omega} \; t
\label{eq:tonshell}
\end{align}
both in $A$ and $A^\dagger$. Time scales here like $t \sim1/g_{\bm k}$ , and $t>t_R$ must be considered in Eq.\ (\ref{eq:tadef}). 
Notice that $t_R$ is a characteristic time of the reservoir and is essentially independent of the operators one considers (in this case $A$ and $A^\dagger$).

\section{Asymptotic reservoir operators}
\label{a:OL}
At (microscopically) large $t$, the part that contains the maximal number of delta functions $\delta(\bar\omega_{\bm k})$ etc.\ dominates (or survives when we consider the weak-coupling limit $g_{\bm k}$ scaling like $1/N\to0$ $\forall{\bm k}$) in each term in the expansion of the asymptotic operator $A^\dagger$ (\ref{eq:Adagger}), while the remaining parts, that depend on the principal values, give minor (or vanishing) contributions owing to the fewer delta functions they contain.  
The same thing happens for the asymptotic operator $A$.  
The asymptotic operators are thus dominated (or effectively given in the weak-coupling limit) by 
\begin{align}
A^\dagger&=2\pi\lambda(ga^\dagger)_{\bm k}\delta(\bar\omega_{\bm k})+2\pi^3\lambda^3(ga^\dagger)_{\bm k_1}\delta(\bar\omega_{\bm k_1})(ga)_{\bm k_2}\delta(\bar\omega_{\bm k_2})(ga^\dagger)_{\bm k_3}\delta(\bar\omega_{\bm k_3})\nonumber\\
&\quad+4\pi^5\lambda^5(ga^\dagger)_{\bm k_1}\delta(\bar\omega_{\bm k_1})(ga)_{\bm k_2}\delta(\bar\omega_{\bm k_2})\cdots(ga^\dagger)_{\bm k_5}\delta(\bar\omega_{\bm k_5})+\cdots ,
\end{align}
where a scale factor $\lambda\propto1/N$ has explicitly been introduced for the coupling $g_{\bm k}\to\lambda g_{\bm k}$.  
It is not difficult to guess the general form in the expansion $2(\lambda\pi)^{2\ell+1}\ell![(ga^\dagger)_{\bm k}\delta(\bar\omega_{\bm k})(ga)_{\bm k'}\delta(\bar\omega_{\bm k'})]^\ell(ga^\dagger)_{\bm k''}\delta(\bar\omega_{\bm k''})$. 
We can therefore regard the asymptotic operators $A^\dagger A$ and $AA^\dagger$ as functions of an operator $\hat x = (ga^\dagger)_{\bm k}\delta(\bar\omega_{\bm k})(ga)_{\bm k'}\delta(\bar\omega_{\bm k'})$,
\begin{equation}
A^\dagger A=f(\hat x),\quad AA^\dagger=f(\hat x+y),
\end{equation}
where $y>0$ is a positive c-number in the weak-coupling limit.   
Since these operators commute with $\rho_a(0)$, they can be simultaneously diagonalized and the normalization conditions, which are nothing but the unitarity relations, are explicitly written as the summations over their respective eigenvalues $E$ and $x$
\begin{equation}
1=\sum_{E\ge0}P_E\sum_{x\ge0}\Bigl({1\over1+f(x)}+{f(x+y)e^{-\beta\Omega}\over1+f(x+y)}\Bigr)\sigma(x)=\sum_{E\ge0}P_Ed_E=\sum_{E\ge0}P_E\sum_{x\ge0}\Bigl({1\over1+f(x+y)}+{f(x)e^{\beta\Omega}\over1+f(x)}\Bigr)\sigma(x),
\end{equation}
where the eigenvalues $x$ are assumed to be distributed over some (positive) range. 
The summation is taken for $d_E=\sum_x\sigma(x)$ terms, with $\sigma(x)$ denoting the degeneracy of eigenvalue $x$, and $d_E$ is the degeneracy of the eigenvalue $P_E$ of the on-shell thermal state $\rho_a^{({\rm on})}(0)$ that is made only of on-shell modes, i.e., $P_E\propto e^{-\beta m\Omega}$, where $m\ge0$ is an integer. 
These relations are rewritten as 
\begin{equation}
\sum_{m\ge0}e^{-\beta m\Omega}\sum_{x\ge0}{f(x)\over1+f(x)}\sigma(x)=e^{-\beta\Omega}\sum_{m\ge0}e^{-\beta m\Omega}\sum_{x\ge0}{f(x+y)\over1+f(x+y)}\sigma(x)=e^{-\beta\Omega}\sum_{m\ge0}e^{-\beta m\Omega}\sum_{x\ge y}{f(x)\over1+f(x)}\sigma(x-y),
\end{equation}
which yields 
\begin{equation}
\sum_{m\ge0}e^{-\beta m\Omega}\sum_{y>x\ge0}{f(x)\over1+f(x)}\sigma(x)=\sum_{m\ge0}e^{-\beta m\Omega}\sum_{x\ge y}{f(x)\over1+f(x)}(e^{-\beta\Omega}\sigma(x-y)-\sigma(x)).
\end{equation}
The relation causes a difficulty, since $f(x)$ is a positive function and the degeneracy generally increases as the eigenvalue increases, that is, $\sigma(x)>\sigma(x-y)$, except near the edge of the spectrum, over which $\sigma(x)$ vanishes. 
A resolution could be found if there exists a unique value $x_m$, eigenvalue of $\hat x$ in the eigenspace characterized by integer $m$, that satisfies ($d_m$ stands for $d_E$ with $E=m\Omega$)
\begin{equation}
\sum_{m\ge0}e^{-\beta m\Omega}{f(x_m)\over1+f(x_m)}d_m=e^{-\beta\Omega}\sum_{m\ge0}e^{-\beta m\Omega}{f(x_m+y)\over1+f(x_m+y)}d_m,
\label{eq:1relation}
\end{equation}
which is equivalent to 
\begin{equation}
\sum_{m\ge0}e^{-\beta m\Omega}\Bigl({1\over1+f(x_m)}+{f(x_m+y)e^{-\beta\Omega}\over1+f(x_m+y)}\Bigr)d_m=\sum_{m\ge0}e^{-\beta m\Omega}d_m=\sum_{m\ge0}e^{-\beta m\Omega}\Bigl({1\over1+f(x_m+y)}+{f(x_m)e^{\beta\Omega}\over1+f(x_m)}\Bigr)d_m.
\label{eq:2relations}
\end{equation}

The above relation (\ref{eq:1relation}) is solved to fix the value of $x_m$.  
Notice first that an apparent solution for function $f$, ${f(x_m)\over1+f(x_m)}=e^{\beta m\Omega}{f(0)\over1+f(0)}$ with $x_m=ym$ gives a vanishing or trivial $f=0$ since $f(0)=0$.  
This implies that one has to take into account of the specific form of $P_E\propto e^{-\beta m\Omega}$. 
We know from the explicit iterative expressions for $A^\dagger A$ and $AA^\dagger$ that $f(x_m)$ and $f(x_m+y)$ are proportional to $x_m$ and $x_m+y$, respectively (this is why $f(0)=0$): therefore, to satisfy the above relation at the lowest order in the expansion, we have
\begin{equation}
\sum_{m\ge0}e^{-\beta m\Omega}x_md_m=e^{-\beta\Omega}\sum_{m\ge0}e^{-\beta m\Omega}(x_m+y)d_m,
\end{equation}
i.e.,
\begin{equation}
\sum_{m\ge0}e^{-\beta m\Omega}x_md_m={ye^{-\beta\Omega}\over1-e^{-\beta\Omega}}\sum_{m\ge0}e^{-\beta m\Omega}d_m=yn\sum_{m\ge0}e^{-\beta m\Omega}d_m,\quad n\equiv{1\over e^{\beta\Omega}-1}.
\end{equation}
This is rewritten as
\begin{equation}
{{\rm Tr}[\hat x\rho_a^{({\rm on})}(0)]\over{\rm Tr}[\rho_a^{({\rm on})}(0)]}=yn={{\rm Tr}[y\overline{a^\dagger a}\rho_a^{({\rm on})}(0)]\over{\rm Tr}[\rho_a^{({\rm on})}(0)]}
\label{eq:lowest}
\end{equation}
where the last equality holds for the averaged number operator for on-shell modes
\begin{equation}
\overline{a^\dagger a}={1\over N_R}\sum_{\bm k,\,\bar\omega_{\bm k}=0}a_{\bm k}^\dagger a_{\bm k},\quad N_R\equiv\sum_{\bm k,\,\bar\omega_{\bm k}=0}.
\end{equation}
The relation would imply $\hat x=y\overline{a^\dagger a}$, which, however, does not satisfy (\ref{eq:1relation}) since $d_m\not=d_{m-1}$, for an arbitrary $f$.
We have to fix $f$ so that the relation is reproduced, which could be done after fixing its c-number counterpart in the following.  

We can easily find the c-number solution to (\ref{eq:lowest})
\begin{equation}
\hat x=y{1\over e^{\beta\Omega}-1}=yn.
\label{eq:cnumber}
\end{equation}
In this case, the operators $A^\dagger A$ and $AA^\dagger$ are replaced by c-numbers $f(yn)$ and $f(yn+y)$. 
Set $f(yn)=yn\tilde f(yn)$, and we have, for $n+1=e^{\beta\Omega}n$, 
\begin{equation}
{\tilde f(yn)\over 1+yn\tilde fy(n)}={\tilde f(yn+y)\over1+(yn+y)\tilde f(yn+y)},
\end{equation}
which is reduced to 
\begin{equation}
{1\over\tilde f(yn)}-{1\over\tilde f(yn+y)}=y.
\end{equation}
The solution is $1/\tilde f(yn)=c-yn$, where $c$ is a constant. We are thus led to the conclusion that in the weak-coupling limit $g_{\bm k}\sim1/N$, unitarity would require the asymptotic operators at $t\sim O(N)$ be replaced by c-numbers
\begin{equation}
A^\dagger A\to {n\over c/y-n},\quad AA^\dagger\to{n+1\over c/y-(n+1)}.
\label{eq:c-sol}
\end{equation}

As a natural expectation, we wonder if the same function with the c-number $yn$ replaced by the operator $\hat x=y\overline{a^\dagger a}$ satisfies the unitarity relations.  
After such a replacement, we observe that
\begin{equation}
{f(\hat x)\over1+f(\hat x)}=y{\overline{a^\dagger a}\over c},\quad{f(\hat x+y)\over1+f(\hat x+y)}=y{\overline{a^\dagger a}+1\over c},
\end{equation}
and therefore their expectation values read
\begin{equation}
\Bigl\langle{f(\hat x)\over1+f(\hat x)}\Bigr\rangle={1\over{\cal Z}^{({\rm on})}}\sum_{m\ge0}e^{-\beta m\Omega}{f(ym)\over1+f(ym)}d_m=y{n\over c}
\end{equation}
and
\begin{equation}
\Bigl\langle{f(\hat x+y)\over1+f(\hat x+y)}\Bigr\rangle={1\over{\cal Z}^{({\rm on})}}\sum_{m\ge0}e^{-\beta m\Omega}{f(ym+y)\over1+f(ym+y))}d_m=y{n+1\over c}.
\end{equation}
Since $n+1=e^{\beta\Omega}n$, we understand that the former expectation value is equal to the latter one multiplied by $e^{-\beta\Omega}$, which is nothing but the relation (\ref{eq:1relation}). 
This means that the unitarity relations have the following operator solution in the weak-coupling limit
\begin{equation}
A^\dagger A=\overline{a^\dagger a}\bigl(c/y-\overline{a^\dagger a}\bigr)^{-1},\quad AA^\dagger=(\overline{a^\dagger a}+1)\bigl(c/y-(\overline{a^\dagger a}+1)\bigr)^{-1}.
\label{eq:opsol}
\end{equation}

We confirm that the expressions (\ref{eq:c-sol}) and (\ref{eq:opsol}) are compatible with the unitarity relations for arbitrary $c/y$ (which expresses a redundancy of the relations, or, alternatively, the fact that the two relations in (\ref{eq:2relations}) are not independent).  
Observe, however, that a constant $c/y$ can be excluded, for it would result in a negative (expectation value of) $A^\dagger A$ for large $n$ (or a large expectation value of $\overline{a^\dagger a}$), for both the c-number solution (\ref{eq:c-sol}) and the operator solution (\ref{eq:opsol}).  
(Even though each term in their series expansions is apparently positive, their convergence is not guaranteed in general.)    
This observation excludes the latter possibility of an operator solution since $c/y$, in this case, is indeed a constant, while there remains the possibility of $c/y>n+1$, that guarantees the positivity in the former case of c-number solution.  
We are thus left with the first possibility: in the weak-coupling limit, the asymptotic operators become c-numbers (\ref{eq:c-sol}). 

Remark that the asymptotic reduced density matrix for the reservoir is nothing but the initial thermal state $\rho_a(0)$ and this conclusion is valid irrespectively of the value of $c/y$, for the relation (\ref{eq:operator})
is satisfied for the c-number solution (\ref{eq:c-sol}) with an arbitrary $c/y$.  
This implies that the reservoir system, initially prepared in a thermal state, returns to the same thermal state after the interaction with the qubit has been turned on.  
We can roughly estimate the time scale necessary for the reservoir to reach the equilibrium as being of order $N$, since the operator $(ga)_{\bm k}\delta(\bar\omega_{\bm k})(ga^\dagger)_{\bm k'}\delta(\bar\omega_{\bm k'})\sim(ga)_{\bm k}t(ga^\dagger)_{\bm k'}t$ brings about a non-vanishing contribution at $t\sim O(N)$ in the weak-coupling limit $g_{\bm k}\sim1/N$.  
Notice that the qubit system is not in equilibrium when the reservoir has relaxed to equilibrium, for the condition $D=0$ is not satisfied for an arbitrary $c/y$.

\bigskip

\section{Vacuum Reservoir}
\label{a:reservoir}
When the reservoir is initially prepared in the vacuum state, i.e., the zero-temperature thermal state, the evolution operators $U_\pm$, etc.\ can be calculated explicitly and the dynamical map is reduced to a simplified form.  
At the same time, however, the treatment at large times has to be reconsidered for we can anticipate that some significant deviations will appear: for instance, the relation $\rho_a(0)a_{\bm k}=e^{\beta\omega_{\bm k}}a_{\bm k}\rho_a(0)$ cannot be valid anymore for $\beta=\infty$ or $\rho_a(0)=|0\rangle\langle0|$ ($a_{\bm k}|0\rangle=0$).  
Since the total excitation number is a conserved quantity and is restricted to zero or one in this case, it is clear that the final state of the reservoir cannot coincide with the initial (vacuum) state, unless the qubit is initially in the ground state (trivial case).  
Consider the case where the qubit is prepared in the excited state and put in contact with the vacuum reservoir.  
The total excitation number is one.
We expect that the qubit will be found in the ground state at large times, with high probability, and thus the reservoir can no longer be in the initial (vacuum) state.  
This is an exceptional case where our conclusion that the equilibrium state is given by the initial thermal state does not apply.  
Incidentally, this is the case on which the previous studies have exclusively focused.

It is important to observe that the operator $U_-$ can be replaced with the identity operator because the annihilation operator that appears at the right-most end in each term of its series expansion (\ref{eq:U-}) 
annihilates the vacuum and this means that $U_-$ which always appears to operate on the vacuum $|0\rangle$ can be set equal to one.  
We can also show that the operator $U_+$ can be evaluated analytically.  
If we look at its series expansion (\ref{eq:U+}), 
we understand that when $U_+$ is applied to the vacuum $|0\rangle$, the creation and annihilation operators $a_{\bm k_{2n-1}}$ and $a_{\bm k_{2n}}^\dagger$, sitting just next to the vacuum, can be replaced by their commutator $\delta_{\bm k_{2n-1},\bm k_{2n}}$.  
The next pair can also be replaced by the Kronecker delta $\delta_{\bm k_{2n-3},\bm k_{2n-2}}$, and so on.  
This means that $U_+$ is given, if applied to the vacuum, by
\begin{align}
&U_+(t)\nonumber\\
&=1-\int_0^tdt_1\int_0^{t_1}dt_2\sum_{\bm k}g_{\bm k}^2e^{-i\bar \omega_{\bm k}(t_1-t_2)}+\int_0^tdt_1\int_0^{t_1}dt_2\int_0^{t_2}dt_3\int_0^{t_3}dt_4\sum_{\bm k_1,\bm k_3}g_{\bm k_1}^2g_{\bm k_3}^2e^{-i\omega_{\bm k_1}(t_1-t_2)}e^{-i\omega_{\bm k_3}(t_3-t_4)}-\cdots\nonumber\\
&=1-\int_0^tdt_1\int_0^{t_1}dt_2\sum_{\bm k}g_{\bm k}^2e^{-i\bar \omega_{\bm k}(t_1-t_2)}U_+(t_2),
\end{align}
which can be solved by a Laplace transformation,
\begin{equation}
U_+(t)=\int{ds\over2\pi i}\tilde U_+(s)e^{st},\qquad \tilde U_+(s)={1\over\ds s+\sum_{\bm k}{g_{\bm k}^2\over s+i\bar\omega_{\bm k}}}={1\over\ds s+\int d\omega{J(\omega)\over s+i\bar\omega}},
\label{eq:U+a}
\end{equation}
where $J(\omega)=\sum_{\bm k}\delta(\omega-\omega_{\bm k})g_{\bm k}^2$ is the spectral density.
Similarly, the evolution operator relevant to the case in which the qubit makes a transition to the ground state reads
\begin{equation}
-i\int_0^t(\bm g\bm a_{t_1}^\dagger)U_+(t_1)=-i\sum_{\bm k}(ga^\dagger)_{\bm k}\int{ds\over2\pi i}{e^{(s+i\bar\omega_{\bm k})t_1}\over s+i\bar\omega_{\bm k}}\Bigl\vert_0^t\tilde U_+(s)=-i\sum_{\bm k}(ga^\dagger)_{\bm k}\int{ds\over2\pi i}\tilde U_+(s){e^{(s+i\bar\omega_{\bm k})t}\over s+i\bar\omega_{\bm k}}.    
\label{eq:flipa}
\end{equation}
The last evolution operator $-i\int_0^t(\bm g\bm a_{t_1})U_-(t_1)$ can be put equal to 0 for it always annihilates the initial vacuum state.

Notice that the above results are exact and satisfy the unitarity relations (\ref{eq:unitarity}) 
when they are applied to the vacuum reservoir. Observe also that one of them is trivially satisfied 
\begin{equation}
U_-^\dagger(t)U_-(t)+\int_0^tU_-^\dagger(t_1)(\bm g\bm a_{t_1}^\dagger)\int_0^t(\bm g\bm a_{t_2})U_-(t_2)=1.
\end{equation}
The other relation is explicitly written as
\begin{align}
&U_+^\dagger(t)U_+(t)+\int_0^tU_+^\dagger(t_1)(\bm g\bm a_{t_1})\int_0^t(\bm g\bm a_{t_2}^\dagger)U_+(t_2)\nonumber\\
&=\int{ds\over2\pi i}\int{ds'\over2\pi i}\Bigl[s+\int d\omega{J(\omega)\over s-i\bar\omega}\Bigr]^{-1}e^{st}\Bigl[1+\sum_{\bm k}{g_{\bm k}^2\over(s-i\bar\omega_{\bm k})(s'+i\bar\omega_{\bm k})}\Bigr]\Bigl[s'+\int d\omega'{J(\omega')\over s'+i\bar\omega'}\Bigr]^{-1}e^{s't}\nonumber\\
&=\int{dsds'\over(2\pi i)^2}{e^{(s+s')t}\over s+s'}\Bigl[s+\int d\omega{J(\omega)\over s-i\bar\omega}\Bigr]^{-1}\Bigl[s+s'+\int d\omega{J(\omega)\over s-i\bar\omega}+\int d\omega'{J(\omega')\over s'+i\bar\omega'}\Bigr]\Bigl[s'+\int d\omega'{J(\omega')\over s'+i\bar\omega'}\Bigr]^{-1}\nonumber\\
&=\int{du\over2\pi i}{e^{ut}\over u}\int {ds\over2\pi i}\Bigl\{\Bigl[s+\int d\omega{J(\omega)\over s-i\bar\omega}\Bigr]^{-1}+\Bigl[s+\int d\omega{J(\omega)\over s+i\bar\omega}\Bigr]^{-1}\Bigr\}=1.
\label{eq:unitaritya}
\end{align}
Observe that while the $u$-integration gives 1, the integration contour over $s$ can be deformed to the right half circle of infinite radius because both terms in the curly brackets have no singularities on the right $s$ half-plane and each term contributes $1/2$.
The unitarity relation is closely related to the probabilities of finding the qubit in the excited or ground states.  
Actually, the term $U_+^\dagger(t)U_+(t)$, which is a c-number, stands for the survival probability of the qubit being found in the excited state. 
The operator $\int_0^tU_+^\dagger(t_1)(\bm g\bm a_{t_1})\int_0^t(\bm g\bm a_{t_2}^\dagger)U_+(t_2)$, on the other hand, becomes the transition probability of the qubit from the excited to the ground states when it is applied to the initial vacuum state.

To extract the asymptotic behavior, we examine $U_+(t)$ in (\ref{eq:U+a}).  
Notice that the integrand has a discontinuity along the imaginary $s$-axis, starting from the branch point at $s=i\Omega$ and extending to $-i\infty$, if the spectral density vanishes for negative $\omega$, i.e., $J(\omega)=0$ for $\omega<0$.
It is not difficult to see that on the imaginary $s$-axis on the first Riemannian sheet, we have a simple pole at $s=is_{\rm I}$ where $s_{\rm I}$ has to satisfy
\begin{equation}
s_{\rm I}=\int d\omega{J(\omega)\over s_{\rm I}+\omega-\Omega}.
\end{equation}
The right-hand side is a well-defined function for $s_{\rm I}>\Omega$ and the above relation has a unique solution at $s_{\rm I}>\Omega$ when the condition
\begin{equation}
\int d\omega{J(\omega)\over\omega}>\Omega    
\end{equation}
is satisfied (strong coupling). 
Incidentally, as is well known, this is the condition for the existence of a discrete eigenstate in the Friedrichs model \cite{Friedrichs1948}.  
For the sake of completeness, we recall here the definition of an ``Ohmic" reservoir:
\begin{equation}
J(\omega)\ \sim \omega^a \quad 
    \begin{cases}
\text{sub-Ohmic}  &  0 < a <1  \\
\text{Ohmic} &  a =1  \\
\text{super-Ohmic}  &   a >1  \\
    \end{cases} .
    \label{eq:ohm}
    \end{equation}
In summary, on the first Riemannian sheet, there is a simple pole on the imaginary $s$-axis when the coupling is strong and a cut connecting $i\Omega$ and $-i\infty$.  
The original integration contour, running from $\epsilon-i\infty$ to $\epsilon+i\infty$, is now deformed into the left-half $s$-plane and the integral over $s$ is evaluated as the sum of contributions from the simple pole and the cut.  
The latter contribution arising from the cut finally decays out for large $t$, while the former, if it exists, yields an oscillating one and never vanishes in the $t\to\infty$ limit. 
That is, we have asymptotically 
\begin{equation}
U_+(t)\sim{\lambda^2e^{i\Omega t}\over(it)^{1+a}}{\Gamma(1+a)\over(\Omega-\int d\omega{J(\omega)\over\omega})^2}+{e^{is_{\rm I}t}\over1+\int d\omega{J(\omega)\over(s_{\rm I}+\omega-\Omega)^2}}.
\end{equation}
The second oscillating term exists only for strong coupling, otherwise $U_+(t)$ decays out like $1/t^{1+a}$ where the spectral function has been assumed to behave like $J(\omega)\sim\lambda^2\omega^a$ ($a>0$) for small $\omega\sim0$.
(Compare with Eq.\ (\ref{eq:ohm}).)

The operator associated with the process where the qubit state flips to the ground state (\ref{eq:flipa}) is asymptotically given by evaluating the residue at $is_{\rm I}$ and the cut contribution.  
Notice that there is no pole at $s=-i\bar\omega_{\bm k}$, which is apparent from the fact that the operator vanishes at $t=0$.  
We have
\begin{align}
&-i\int_0^t(\bm g\bm a_{t_1}^\dagger)U_+(t_1)\nonumber\\
&=-i\sum_{\bm k}(ga^\dagger)_{\bm k}\biggl\{\int_{-\infty}^\Omega{ds\over2\pi}{e^{i(s+\bar\omega_{\bm k})t}\over i(s+\bar\omega_{\bm k})}[\tilde U_+(\epsilon+is)-\tilde U_+(-\epsilon+is)]+{e^{i(s_{\rm I}+\bar\omega_{\bm k})t}\over i(s_{\rm I}+\bar\omega_{\bm k})}{1\over1+\int d\omega{J(\omega)\over(s_{\rm I}+\bar\omega)^2}}\biggr\}\nonumber\\
&\sim-i\sum_{\bm k}(ga^\dagger)_{\bm k}\int_{-\infty}^\infty{dx\over2\pi i}{e^{ix}\over x}[\tilde U_+(\epsilon-i\bar\omega_{\bm k})-\tilde U_+(-\epsilon-i\bar\omega_{\bm k})]-\sum_{\bm k}(ga^\dagger)_{\bm k}{e^{i(s_{\rm I}+\bar\omega_{\bm k})t}\over s_{\rm I}+\bar\omega_{\bm k}}{1\over1+\int d\omega{J(\omega)\over(s_{\rm I}+\bar\omega)^2}}\nonumber\\
&=-{i\over2}\sum_{\bm k}(ga^\dagger)_{\bm k}[\tilde U_+(\epsilon-i\bar\omega_{\bm k})-\tilde U_+(-\epsilon-i\bar\omega_{\bm k})]-\sum_{\bm k}(ga^\dagger)_{\bm k}{e^{i(s_{\rm I}+\bar\omega_{\bm k})t}\over s_{\rm I}+\bar\omega_{\bm k}}{1\over1+\int d\omega{J(\omega)\over(s_{\rm I}+\bar\omega)^2}}.
\end{align}

The unitarity relation (\ref{eq:unitarity}) 
at large times implies
\begin{equation}
{1\over1+\int d\omega{J(\omega)\over(s_{\rm I}+\bar\omega)^2}}+{1\over4}\int d\omega J(\omega)\bigl|\tilde U_+(\epsilon-i\bar\omega)-\tilde U_+(-\epsilon-i\bar\omega)\bigr|^2=1
\label{eq:unitarity1a}
\end{equation}
for strong coupling, while the first term would be absent for weak coupling. 
The equality, which is required by the unitarity, actually holds.  
Since $\tilde U_+(s)$ has a simple pole (at $s=is_{\rm I}$) and the cut (along the imaginary $s$-axis between $i\Omega$ and $-i\infty$), and behaves like $1/s$ at infinity, we have an identity relation
\begin{equation}
\lim_{R\to\infty}\oint_{s=Re^{i\theta}}{ds\over2\pi i}\tilde U_+(s)=1=\int_{-\infty}^{\Omega}{ds\over2\pi}[\tilde U_+(\epsilon+is)-\tilde U_+(-\epsilon+is)]+{1\over1+\int d\omega{J(\omega)\over(s_{\rm I}+\omega)^2}}.
\label{eq:identitya}
\end{equation}
Notice that the last integral in (\ref{eq:unitarity1a}) is rewritten as
\begin{align}
&\int d\omega J(\omega)|\tilde U_+(\epsilon-i\bar\omega)-\tilde U_+(-\epsilon-i\bar\omega)|^2\nonumber\\
&=\int d\omega J(\omega)\Bigl[{1\over iR+\pi J(\omega)}{1\over-iR+\pi J(\omega)}-{1\over iR+\pi J(\omega)}{{1\over-iR-\pi J(\omega)}}\nonumber\\
&\qquad\qquad\qquad
-{1\over iR-\pi J(\omega)}{1\over iR+\pi J(\omega)}+{1\over iR-\pi J(\omega)}{1\over-iR-\pi J(\omega)}\Bigr]\nonumber\\
&=\int{d\omega\over2\pi}\Bigl[{1\over iR+\pi J(\omega)}+{1\over-iR+\pi J(\omega)}+{1\over iR+\pi J(\omega)}-{{1\over-iR-\pi J(\omega)}}\nonumber\\
&\qquad\qquad
-{1\over iR-\pi J(\omega)}+{1\over iR+\pi J(\omega)}-{1\over iR-\pi J(\omega)}-{1\over-iR-\pi J(\omega)}\Bigr]\nonumber\\
&=4\int{d\omega\over2\pi}\Bigl[{1\over iR+\pi J(\omega)}+{1\over-iR+\pi J(\omega)}\Bigr]\nonumber\\
&=4\int{d\omega\over2\pi}[\tilde U_+(\epsilon-i\bar\omega)-\tilde U_+(-\epsilon-i\bar\omega)]=4\int_{-\infty}^{\Omega}{ds\over2\pi}[\tilde U_+(\epsilon+is)-\tilde U_+(-\epsilon+is)],
\end{align}
which, together with (\ref{eq:identitya}), implies that the equality (\ref{eq:unitarity1a}) holds, where we have set
\begin{equation}
\tilde U_+(\epsilon-i\bar\omega)={1\over-iR+\pi J(\omega)},\quad\tilde U_+(-\epsilon-i\bar\omega)={1\over-iR-\pi J(\omega)},\quad R\equiv\bar\omega-{\rm P}\int d\omega'{J(\omega')\over\bar\omega-\bar\omega'}.
\end{equation}

We see that if the reservoir interaction with the qubit is strong enough, there remains a finite probability that the qubit and the reservoir remain in the initial state and a complete relaxation never occurs for the qubit. 
Such a finite probability of the final qubit being in the initial excited state is due to the existence of a discrete eigenstate (bound state)  of the total Hamiltonian.  
On the other hand, if the interaction is not strong, the qubit completely decays to the ground state and the reservoir is in an excited state, not in the vacuum.  
This system is thus considered to be an exceptional case to which the conclusion of the main text does not apply.

\end{document}